# Electrical Tuning of Plasmonic Conducting Polymer Nanoantennas

Akchheta Karki[1], Giancarlo Cincotti[1], Shangzhi Chen[1], Chuanfei Wang[1], Vallery Stanishev[2,3], Vanya Darakchieva[2,3], Mats Fahlman[1], Magnus P. Jonsson[1]★

[1]Laboratory of Organic Electronics, Department of Science and Technology (ITN), Linköping University, SE-601 74 Sweden
[2]Terahertz Materials Analysis Center (THeMAC), Department of Physics, Chemistry and Biology (IFM), Linköping University, Linköping, SE-581 83 Sweden
[3]Center for III-Nitride Technology, C3NiT-Janzèn, Department of Physics, Chemistry and Biology (IFM), Linköping University, Linköping, SE-581 83 Sweden
★Correspondence: magnus.jonsson@liu.se

**Nanostructures of conventional metals offer manipulation of light at the nanoscale but are limited to static behavior due to their fixed material properties. To develop the next frontier of dynamic nanooptics and metasurfaces, we utilize the redox-tunable optical properties of conducting polymers, which were recently shown to be capable of sustaining plasmons in their most conducting oxidized state. Using nanodisks of poly(3,4-ethylenedioxythiophene:sulfate) (PEDOT:Sulf) as a model system, we present the first electrically tunable conducting polymer nanooptical antennas. In addition to repeated on/off switching of the polymeric nanoantennas, we demonstrate the possibility for gradual electrical tuning of their nanooptical response, which was found to be related to the modulation of both density and mobility of the mobile polaronic charge carriers in the polymer. The presented concept takes important steps towards electrically tunable metasurfaces with truly dynamic optical nanoantenna pixels, with not only varying farfield but also tunable nearfield. The work paves the way for applications ranging from tunable flat metaoptics to adaptable smart windows.**

# 1. Introduction

Metal nanostructures can be used as optical nanoantennas by converting free-space optical radiation into collective charge oscillations called plasmons[1]. Due to their ability to control light at the nanoscale, such systems have been utilized in areas including energy conversion[2–4], biosensing[5–9], display technologies[10–13], ultrathin optical components[14–18], and many others. However, light-matter interactions with conventional metal nanostructures are limited as it has proven unduly challenging to modify their static properties after fabrication[19]. To circumvent this major issue and open avenues to dynamically control light at the nanoscale, research is shifting towards dynamic systems with tunable properties, for example, based on phase change materials[20–24], doped metal oxide nanocrystals[25], and graphene[26–28]. Motivated by an exceptionally large redox-tunability[29], we recently introduced conducting polymers as a new platform for dynamic plasmonics[30]. We showed that nanodisks of the highly conducting polymer poly(3,4-ethylenedioxythiophene:sulfate) (PEDOT:Sulf) can function as dynamic plasmonic nanoantennas, with plasmons originating from the highly mobile and large density of polaronic charge carriers ($2.6 \times 10^{21}$ cm$^{-3}$, determined by ellipsometry) in the polymer network[30]. Excitingly, these nanoantennas could be completely switched on and off by chemical tuning of the polymer's redox state, which dramatically modulates the conductivity and optical properties of the material[30]. However, the tuning process was based on exposure to gases and liquids while future systems will require more convenient and faster electrical tuning.

In this work, we take the next crucial step in the study of conducting polymer plasmonics and present the first electrically tunable conducting polymer optical nanoantennas. Using PEDOT:Sulf nanodisks as a model system, we demonstrate repeatable and complete on/off switching by electrochemically switching the polymer between its highly conducting plasmonic oxidized state and a lower conducting dielectric reduced state. In addition to complete and fast (seconds timescale) responsive on/off switching, we demonstrate gradual tuning of the plasmonic response by gradual modulation of the polymer redox state. Complementary analytical calculations and simulations suggest that the tunable response originates from modulation of both density and mobility of the mobile polaronic charge carriers in the polymer. The combination of gradual tuning of the nanoantennas and the ability to turn them completely on and off enables many future applications in the direction of dynamic nanooptics and metasurfaces, ranging from smart windows with controllable plasmon-induced heating to dynamic ultrathin optical components for beam steering or lensing with tunable focal length[31,32].

# 2. Results and Discussions

Fig. 1a shows the chemical structures of PEDOT:Sulf in its oxidized (plasmonic) and reduced (dielectric) states, which form the basis for its use as a dynamic plasmonic material. Fig. 1b compares the in-plane permittivity of PEDOT:Sulf in the two states, indeed showing a clear transition between metallic (negative real permittivity) and dielectric (positive real permittivity) response. The data for the oxidized state was obtained from our previous work[30] and we measured the permittivity of chemically reduced PEDOT films (Supplementary Fig. 1, Fig. 2, and Supplementary Table 1 contain the raw ellipsometry data, fitting parameters, and the out-of-plane permittivity). The oxidized PEDOT:Sulf film provides negative real permittivity and a lower magnitude imaginary permittivity in the spectral region from 0.8–3.6 μm, which was defined as the plasmonic regime[30]. Contrastingly, the reduced material instead largely shows positive permittivity in the same spectral region, presenting an opportunity to tune PEDOT:Sulf based systems between its plasmonic and dielectric behavior.

Indeed, simulations using experimentally obtained material properties for the oxidized and reduced states demonstrate the potential for tuning both farfield and nearfield responses of PEDOT:Sulf nanoantennas. The blue curve in Fig. 1c shows the optical extinction of an oxidized (plasmonic) PEDOT:Sulf nanodisk array, exhibiting a clear resonance peak at a wavelength of around 1500 nm. By contrast, this resonance peak is not present for the same nanodisk array made of the chemically reduced (dielectric) PEDOT (black curve). Instead, there are two peaks at wavelengths of ~1000 nm and ~630 nm, which are attributed to the low oxidation and neutral states of PEDOT, respectively[33]. The simulated data is comparable to the experimentally obtained extinction spectra of oxidized and reduced PEDOT:Sulf nanodisks, as presented in Supplementary Fig. 3. Fig. 1d compares the optical near-field profiles of the oxidized and reduced states of a nanodisk (at the resonance wavelength of ~1500 nm), demonstrating that the intensity of the enhanced electric fields around the nanoantenna is significantly higher in the oxidized state, thereby confirming the expected tunability also of the nearfield nanoantenna response.

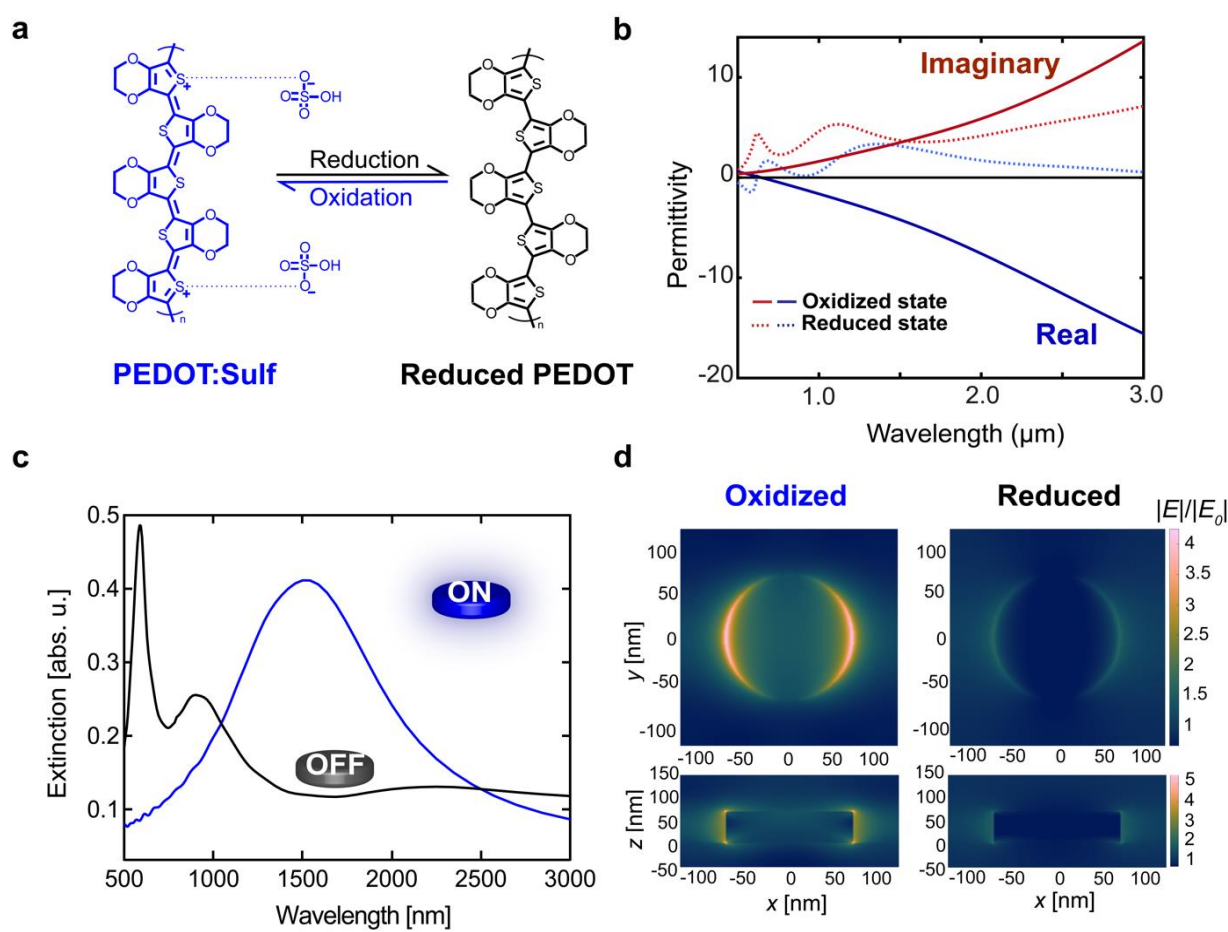

**Fig. 1: Concept of electrically tunable conducting polymer nanoantennas. a**, Molecular structures of PEDOT:Sulf in its oxidized (plasmonic) and reduced (dielectric) states. **b**, In-plane permittivity dispersion of PEDOT:Sulf in its oxidized (blue curve, real part; red curve, imaginary part) and reduced (dashed blue curve, real part; dashed red curve, imaginary part) states. The spectral range between 0.8 to 3.6 μm is defined as a plasmonic regime where the real permittivity is below zero and its magnitude is larger than the imaginary component[30]. **c,d**, Simulated extinction (c) and nearfield (d) response for nanodisk arrays (diameter=145 nm, thickness=65 nm, period=600 nm) at the resonance peak wavelength of ~1500 nm, based on fully oxidized (blue) and chemically reduced (black) PEDOT:Sulf. In (d), top panels are top views while bottom panels are cross sections of the nanodisks along the direction of the incident

polarization, which is along the *x*-axis. *x*–*y* in-plane direction is 2 nm above the nanodisk and *x*–*z* cross-section is through the center of the nanodisk. The color scale bars show the electric field strength relative to the incident light ($|E|/|E_0|$).

To experimentally confirm excitation of plasmons in PEDOT:Sulf polymer nanodisks and demonstrate tuning of their plasmonic response, we fabricated sparse arrays of nanodisks of different dimensions on sputtered ITO substrates using colloidal lithography (detailed schematic of the polymerization and colloidal lithography process can be found in Supplementary Fig. 4). Atomic force microscopy (AFM) images reveal successful fabrication of large areas of nanodisks (Fig. 2a(i)-c(i)). Height and width section analyses of individual nanodisks from AFM images (Supplementary Fig. 5-7) were used to extract the representative heights and diameters, as presented in Fig. 2a(ii)-c(ii). Our previous work established that decreasing the diameter of the nanodisks could blue shift the resonance peaks, whereby resonance peaks down to 1800 nm were attained for around 30 nm thick PEDOT[30]. In this work, we instead limited the variation in nanodisk diameter (149 nm-196 nm) and increased the thickness of the nanodisks, leading to significantly blue shifted peaks, with lowest resonance wavelength of around 1270 nm. By sequentially increasing the nanodisk thickness from ~30 nm (based on one layer of vapor phase polymerized (VPP) PEDOT:Sulf), to ~60 nm (two layers of VPP PEDOT:Sulf), and finally ~90 nm (three layers of VPP PEDOT:Sulf), we achieved substantial blue shifts in the experimental resonance peaks, from ~1950 nm down to ~1270 nm (Fig. 2a(iii)-c(iii)) for around 150 nm in diameter nanodisks. The small variations in nanodisk diameter are related to separate optimization of the polymer etching time for the three different nanodisk conditions to account for the differences in film thickness. The sometimes negative extinction values at longer wavelengths are attributed to variations in the extinction of the sputtered glass/ITO reference substrates (see Supplementary Fig. 8 for more details). The experimental results match the simulated results in terms of both peak wavelengths and peak widths (Fig. 2a(iv)-c(iv)). Slight discrepancies can be attributed to geometrical variations and imperfections of the fabricated nanodisks. The large resonance blue-shifts, using only minor modifications in geometry, is promising for achieving resonances also in the visible spectral range, especially in combination with further materials optimization.

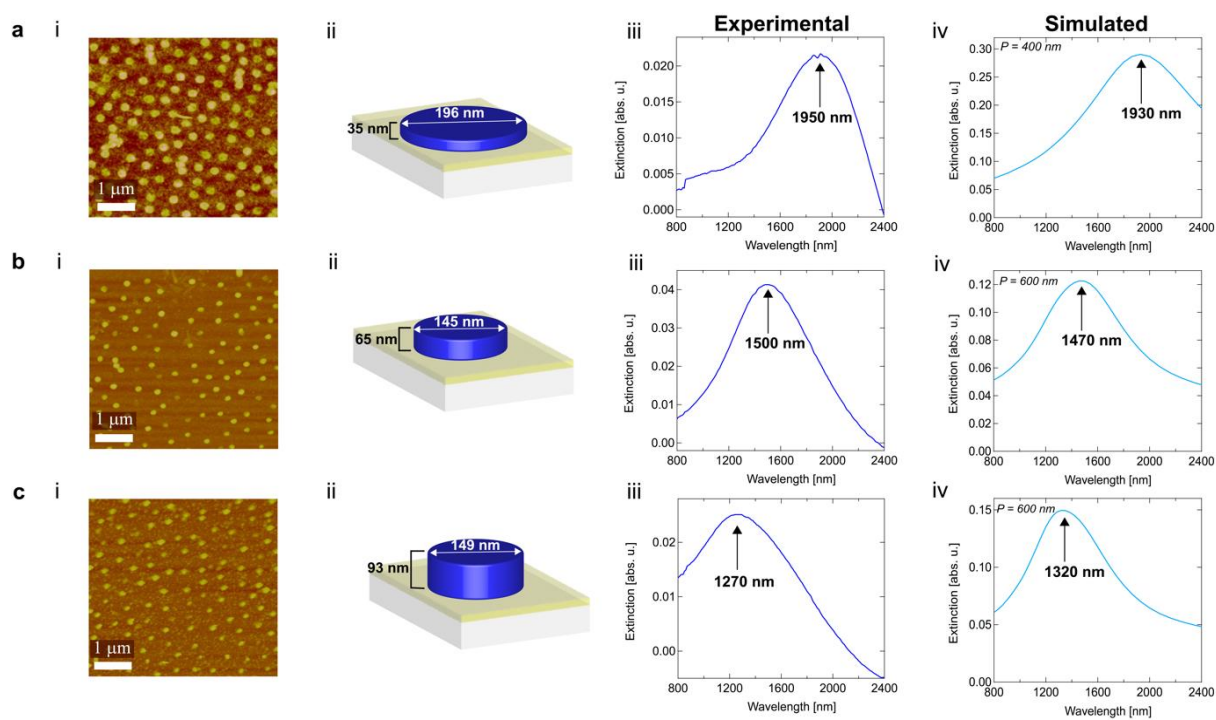

**Fig. 2: Resonance-tuning by nanodisk geometry. a-c,** Nanodisks made using one (a), two (b), and three (c) PEDOT:Sulf layers to control the thickness, with (i) AFM images (ii) schematic of nanodisks with extracted dimensions from AFM height and width analyses. (iii) Experimental extinction spectra and (iv) simulated extinction spectra for the nanodisks of the corresponding dimensions. The different values of *P* indicated in each panel in (iv) correspond to the array period in the periodic simulations.

Next, we demonstrate electrical switching of the conducting polymer nanoantenna surfaces. We here utilize that the redox state of conducting polymers can be modulated electrochemically[34–38]. Fig. 3a shows the basic device structure, based on PEDOT:Sulf nanodisks on a glass/ITO substrate, coated by an ion gel followed by a second ITO/glass substrate. The two ITO layers act as electrodes and are separated by the ion gel containing the ionic liquid 1-Ethyl-3-methylimidazolium bis(trifluoromethylsulfonyl)imide ([EMIM$^+$][TFSI$^-$]) in a mixture of the co-polymer poly(vinylidene flouride) (PVDF-HFP) in acetone. Figure 3b shows the chemical structures of the components in the ion gel, which provide high transparency in the visible to mid-IR ranges (see Supplementary Fig. 9a). It is worth noting that applying the ion gel on top of the nanodisks red-shifted the plasmonic resonance (Supplementary Fig. 9b), as expected due to the increase in refractive index of the surrounding medium from 1 ($n_{air} = 1$) to that of the ion gel ($n_{ion-gel} \approx 1.42$)[39,40].

Fig. 3c shows extinction spectra of a complete device, demonstrating that the resonance peak observed in the oxidized plasmonic state (at 0V, blue curve) could be completely suppressed by applying a positive bias of 5V from the top ITO electrode, which effectively turns off the nanoantennas (grey curve). The nanoantennas could then be switched on again by applying a 0V potential and thereby re-oxidizing the material to make it plasmonic again (dashed blue curve). To better understand the switching behavior, we note that the PEDOT:Sulf nanodisks are in their oxidized state at 0 V, with high density mobile polaronic charge carriers giving rise to the plasmonic response. At this state, the mobile positive polaronic charges in the PEDOT:Sulf nanodisks are being compensated by the sulfate counterions ($HSO_4^-$) as depicted in Fig. 1a. Upon applying a positive bias from the top ITO electrode via the ion gel, cations (EMIM$^+$) from the ion gel get injected into the polymer nanodisks, which are compensated by the $HSO_4^-$ anions[37]. The presence of EMIM$^+$ in the PEDOT:Sulf film after applying a positive bias was confirmed by X-ray photoelectron spectroscopy (XPS), showing clear nitrogen signals due to the N and N+ signals present in EMIM$^+$ (Supplementary Fig. 10a). Additionally, the sulfur signal from $HSO_4^-$ moved towards higher binding energies after switching, which could be an indication of the sulfur from $HSO_4^-$ coordinating with $N^+$ from the cation EMIM$^+$ rather than with the positive polaronic charge carriers in PEDOT:Sulf (as in the oxidized state) (Supplementary Fig. 10b). Consequently, applying a positive bias decreases the density of holes in the PEDOT:Sulf due to the lack of coordination between $HSO_4^-$ and the positive polaronic charge carriers in the polymer film[37]. As a result, the polymer nanoantennas reach an off state where the number of mobile polaronic charge carriers are so low that the film no longer exhibits plasmonic behavior, as shown by the grey curve in Fig. 3c.

To further understand the dynamics of the switching process, we monitored the optical extinction (normalized from 0 to 1 relative to the first switching cycle) at the plasmonic peak position of around 1800 nm in real time while repeatedly switching the bias between 0V to 5V (Fig. 3d). The results show reversible transitions between the on (0V) and off (5V) states for at least 50 switching cycles (50 minutes) (see results for all cycles in Supplementary Fig. 11), albeit with some reduction in the extinction values after many cycles. The inset in Fig. 3d shows that it took only a few seconds for the nanodisk extinction signal to reach the off state after applying the bias, and a further 20-30s for the values to fully stabilize. Likewise, it took about 20-30s for the extinction values to stabilize after turning the nanoantennas on again (i.e. removing the bias). These values, which are based on devices not optimized for speed, are in line with typical conducting polymer electrochemical devices with similar solid ion gel-based electrolytes[38,41,42]. Because the switching speed is strongly dependent on the mobility of the ions in the solid ion gel, there is potential for further improvements by optimizing the electrolyte to obtain higher ionic mobility[43,37,41]. Furthermore, recent research also showed that modified device configurations can boost the switching speed of conducting polymer devices even to video rates[44].

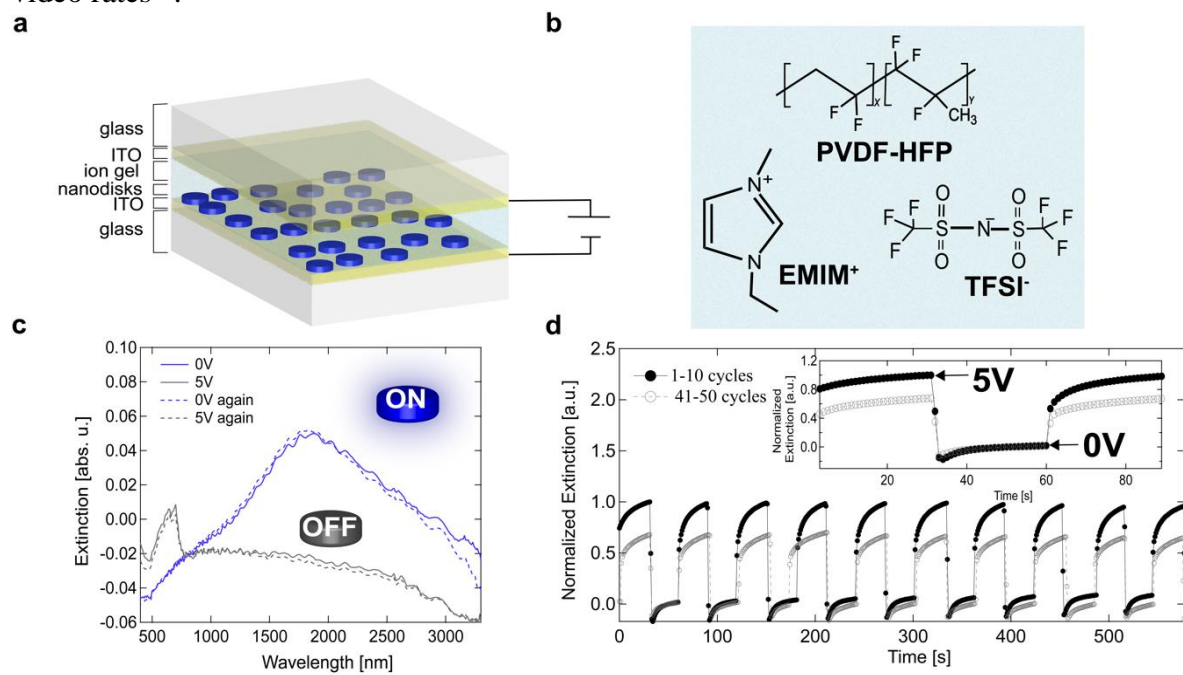

**Fig. 3: Reversible electrical on/off switching of PEDOT:Sulf nanoantennas. a**, Schematic illustration of the electrically tunable device, based on conducting polymer nanodisks on an ITO electrode, electrochemically modulated via a transparent ion gel and a second top ITO electrode. **b**, Chemical structures of the components in the ion gel. **c**, Extinction spectra for a device with 145 nm in diameter and 65 nm thick nanodisks at 0V and 5V, with dashed lines showing excellent reversibility upon repeated switching. The small undulations are attributed to optical interference in the thin film device structure due to the sandwiched ion gel, as also observed without the nanodisks (Supplementary Fig. 9a) and previously reported for similar ion gel components and device structures as used in this work[38]. **d**, Normalized extinction at the plasmonic peak position of around 1800 nm over time during multiple (50) switching cycles, with the inset showing results for the first and 41$^{st}$ cycles.

We will now demonstrate that the plasmonic response of conducting polymer nanoantennas can be tuned gradually, in addition to complete electrical on/off switching. Fig. 4a shows experimental extinction spectra at different biases applied from the top ITO electrode, for a device with 145 nm in diameter and 65 nm thick PEDOT:Sulf nanodisks. The results show gradual suppression of the nanoantenna response accompanied by a small red-shift of the extinction peak position. Similar results were obtained for PEDOT:Sulf nanodisks with other thicknesses (Supplementary Fig. 12).

During the gradual tuning process, the PEDOT:Sulf nanodisks evolve towards a reduced state with increasing applied bias, as confirmed by the appearance of a peak at ~1000 nm at 3V related to the lower oxidation state of PEDOT[33]. Upon further increasing the bias to 5V, a peak at ~630 nm originates from the reduced neutral PEDOT[33]. In this reduced state, any signs of plasmons originating from the mobile polaronic charges have been diminished as a result of polymer reduction. Control measurements using the same device configuration as used for the nanodisks but with a non-structured PEDOT:Sulf film confirm that the polymer is gradually reduced with an increasing bias (Supplementary Fig. 13), including reduction of the IR absorbance and emergence of the neutral (~600 nm) and polaronic (~1000 nm) peaks. While reduction of the PEDOT:Sulf nanodisks was not clearly visible by the eye, the reduced PEDOT:Sulf film also took on its traditional deeper blue coloration (Supplementary Fig. 13b)[45–48].

To better understand the dynamic nanooptical response of the PEDOT nanoantennas and the relation to changes in the polaronic charge transport at different biases, we calculated their extinction cross section ($\sigma(\lambda)$) using dipolar polarizability theory and by varying the in-plane polymer permittivity ($\varepsilon(\lambda)$) via changes in the charge carrier mobility and/or concentration. Treating the nanodisks as oblate spheroids with diameter $D$ and thickness $t$, the polarizability ($\alpha$) in the quasi-static limit ($D \ll \lambda$) is given by[49]:

$$\alpha(\lambda) = V \frac{\varepsilon(\lambda) - \varepsilon_s}{\varepsilon_s + L[\varepsilon(\lambda) - \varepsilon_s]} \quad (1)$$

where $V$ is the volume of the spheroid, $L$ is a geometrical factor that depends on spheroid aspect ratio, and $\varepsilon_s$ is the permittivity of the surrounding medium (set to 2.13 to resemble the effective surrounding permittivity for disks surrounded by the ion gel on a glass substrate). As larger disks require correction for finite wavelength effects, we further correct the polarizability using[50,51]:

$$\alpha'(\lambda) = \alpha(\lambda)\left[1 - \frac{k^2}{2\pi D}\alpha(\lambda) - i\frac{k^3}{6\pi}\alpha(\lambda)\right] \quad (2)$$

where $k$ is the wavenumber of the incident light and $i$ is the imaginary number. The extinction cross-section $\sigma(\lambda)$ of the nanodisks/spheroids can now be calculated as:

$$\sigma(\lambda) = k\mathrm{Im}[\alpha'(\lambda)] \quad (3)$$

The black curves in Fig. 4b-d show the calculated extinction for a fully oxidized PEDOT:Sulf nanodisk, with good agreement with the experimental results (Fig. 4a) in terms of both resonance position and peak width. $\varepsilon(\lambda)$ was here described using a Drude-Lorentz model as detailed in previous reports[30,52]:

$$\varepsilon(\lambda) = \varepsilon_\infty - \frac{\omega_p^2 \tau}{\omega^2 \tau + i\omega} - \sum_k \frac{A_j}{\omega^2 - \omega_j^2 + i\omega\gamma_j} \quad (4)$$

where $\varepsilon_\infty$ is the high frequency permittivity offset (beyond the measurement range). The second term is the Drude component, for which $\tau$ is the momentum-averaged scattering time, $\omega$ is the angular frequency, and $\omega_p$ is the plasma frequency. The last part of equation 4 describes Lorentz oscillators representing other features including molecular resonances and anomalous optical conductivity behavior, where $A_j$, $\omega_j$, and $\gamma_j$ are amplitude, resonance angular frequency, and broadening for the $j^{th}$ Lorentz oscillator, respectively. Importantly, $\omega_p$ and $\tau$ are directly related to the free charge carrier density ($n$) and DC mobility ($\mu$) via equations S1 and S2 (Supplementary Note A). We can therefore calculate the extinction for the same nanodisks but with material permittivity corresponding to different $n$ and/or $\mu$.

Fig. 4b shows the calculated nanodisk extinction spectra upon artificially decreasing the charge carrier density of the polymer via the Drude term of equation 4. (In the Supplementary Information we provide results of also analogously modifying the amplitudes ($A_j$) and broadenings ($\gamma_j$) of the Lorentz oscillator terms ($\sum_k \frac{A_j}{\omega^2 - \omega_j^2 + i\omega\gamma_j}$) (Supplementary Fig. 14).)

In line with the experiments, the resonance peak and nanooptical responses are highly suppressed with decreasing $n$. Our control measurements also confirm a significant drop in the charge carrier density from 100% to around 10% with applied biases (obtained from the integrated current/volume over time upon reduction of a polymer film, see Supplementary Fig. 15 and Supplementary Table 2 for details). However, the analytical calculations also suggest a significant red-shift of the resonance peak position with decreasing $n$ while the experimentally measured extinction spectra show more modest red-shifts upon applying biases. Similar large calculated red-shifts were also obtained if modulating both the Drude and the Lorentz oscillator terms in equation 4 (Supplementary Fig. 14). This discrepancy suggests that decreased charge carrier density alone does not account for the changes observed in the experimental extinction spectra. Interestingly, the analytical calculations for decreasing $\mu$ instead of $n$ (Fig. 4c) do not predict a red-shift but rather a blue-shift if modulating the Drude-term of equation 4 (and hardly any shift in peak position if also modulating the Lorentz oscillators, Supplementary Fig. 14). Reduction in the charge carrier mobility as the system reaches a reduced state may therefore complement the reduction in carrier density and contribute to the lower experimental red-shift. To experimentally investigate if the mobility decreases upon reduction, we measured the conductivity ($\sigma$) of a chemically reduced PEDOT:Sulf film and estimated the mobility using $\sigma = qn\mu$, where $q$ is the elementary charge. Using $n$ obtained for the electrochemically reduced film, the calculation suggests that the mobility may drop down to ~1.5 % of the mobility of the fully oxidized plasmonic film (see Supplementary Note B). Although this value may be underestimated considering that the carrier density is likely suppressed more upon chemical rather than electrochemical doping, the experiments corroborate that the tuning of the nanodisk antennas involves reduction of both carrier density and mobility. Indeed, drops in $\mu$ upon reduction has been reported previously for similar polymers, as a result of reversible changes in microstructure such as degree of order and lamellar packing distance[53,54]. Calculated effects of simultaneously changing $n$ and $\mu$ can be seen in Fig. 4d as an overall reduction in the extinction signals with a slight red-shift. Therefore, comparison of the analytical calculations with the experimental results in Fig. 4a suggests that the experimental device undergoes a combined reduction of carrier density and mobility. Future work may also investigate possible influence of other variations upon reduction, including the emergence of the neutral (~600 nm) and polaronic (~1000 nm) peaks, which may also reduce the red-shift.

Finally, we investigate effects of redox-tuning on the optical nearfields by numerically simulating the response for a single nanodisk while jointly varying both $n$ and $\mu$. For the fully oxidized nanodisk, we used the anisotropic complex permittivity of the original oxidized PEDOT:Sulf and then varied both the in-plane and out-of-plane permittivity via $n$ and $\mu$ using the approach described above. As seen in Fig. 4e-h, gradual reductions in $n$ and $\mu$ from 100% (oxidized), to 70%, 30%, and 1% (reduced), greatly decreases the strength of the plasmon-enhanced fields on the opposite edges of the nanodisk. In its oxidized (plasmonic) state, the nanodisk exhibits a clear dipolar nearfield profile with enhanced fields, which then gradually decreases and disappears as $n$ and $\mu$ decrease. The corresponding simulated extinction profiles are also consistent with this finding (Supplementary Fig. 16).

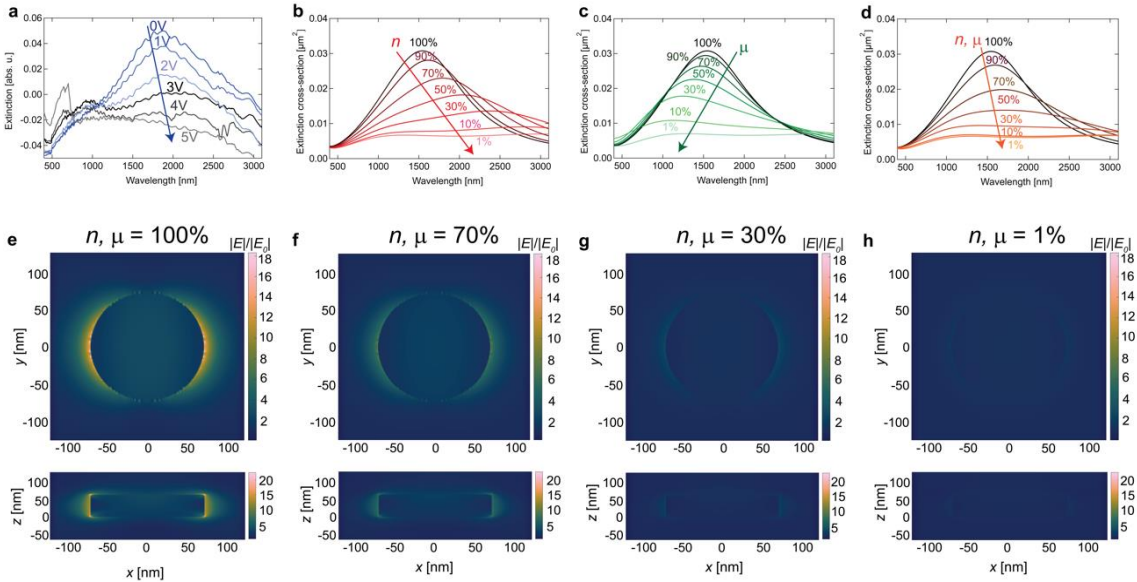

**Fig. 4: Gradual electrical tuning of conducting polymer nanoantennas. a**, Experimental extinction results at different biases, showing the capability of gradual tuning of the nanoantenna response for a device with 145 nm in diameter and 65 nm thick PEDOT:Sulf nanodisks. **b-d**, Calculated extinction based on dipolar polarizability theory of the nanodisks treated as oblate spheroids (same dimensions as in (a)), with material properties varying from the fully oxidized material by gradually decreasing (b) carrier density, (c) carrier mobility, (d) both carrier density and mobility. **e-h**, Simulated nearfields for a single nanodisk on glass (same dimensions as in (a)) at the resonance peak with jointly varying carrier density and mobility, from (e) 100% (fully oxidized), (f) reduction to 70% of both carrier density and mobility, (g) reduction to 30%, and (h) reduction to 1% of the original carrier density and mobility. The color scale bars show the electric field strength relative to the incident light ($|E|/|E_0|$). $x$–$y$ in-plane direction is through the middle of the nanodisk height ($z=32$ nm) and $x$–$z$ cross-section is through the center of the nanodisk.

## 3. Conclusions

To summarize, we have demonstrated the first electrically tunable conducting polymer optical nanoantennas. The polymeric nanoantennas provide repeated and fast (seconds timescale) on/off switching as well as gradual tuning of their nanooptical response. The behavior stems from the ability to tune the polymer material between being optically metallic and dielectric. Complementary analytical calculations and control experiments show that the redox-tuning involves gradual modulation of both density and mobility of the polaronic charge carriers in the polymer. Further simulations reveal the possibility to modulate also the optical nearfield response of the nanoantennas. The concept takes important steps towards electrically tunable metasurfaces based on optical nanoantenna pixels with electrically tunable behavior.

## 4. Methods

*Thin film deposition*
Glass substrates (7.2 × 2.5 cm$^2$) were cleaned by sequential ultrasonication in a hellmanex solution in deionized (DI) water, DI water, acetone, and isopropanol for 15 minutes, respectively, followed by air drying with nitrogen. The glass substrates were then put inside a Vaksis PVD MIDAS 3M thin film coating system (VK-1902) to sputter 50 nm of ITO (sputtering conditions: 43 % Ar, 25 °C, 32 W power). Following the sputtering process, the ITO/glass substrates were diced in half (3.6 × 2.5 cm$^2$) and again cleaned by sequential ultrasonication in a hellmanex solution in DI water, DI water, acetone, and isopropanol for 15 minutes, respectively, followed by air drying with nitrogen. The films were finally treated with oxygen plasma at 100 W for 5 mins before beginning the film deposition process. For the film deposition, PEDOT:trifluromethanesulfonate (PEDOT:OTF) thin films were first prepared as precursors of PEDOT:Sulf films. PEDOT:OTF thin films were deposited via vapor phase polymerization as reported in the literature[55]. The oxidant solution for EDOT polymerization was prepared by mixing 0.03 g of iron(III) trifluoromethanesulfonate 90% (Fe(OTf)$_3$ (Sigma Aldrich), 0.2 g of the triblock co-polymer poly(ethylene glycol)-block-poly(propylene glycol)-block-poly(ethylene glycol) (average $M_n \approx 5,800$ (Sigma-Aldrich)) and 0.8 g of 99.5% ethanol (Solveco). Oxidant films were deposited by spin coating at 1500 rpm for 22 s onto precleaned ITO substrates. After 30 s of baking on a hotplate at 70 °C, the samples were transferred inside a heated vacuum desiccator (Vacuo-temp (SELECTA)) for the vapor phase polymerization (VPP) of PEDOT:OTf. 100 μL of EDOT (142.18 g mol$^{-1}$ (Sigma-Aldrich)) droplets were drop cast onto two glass substrates on a hotplate at 60 °C in the desiccator to ensure its evaporation during polymerization. After 45 min of polymerization at a static pressure of 70 mbar, the samples were taken out from the desiccator and submerged inside a petri-dish containing ethanol for 2 hours. For two times and three times VPP, 45 min of polymerization was followed by soaking in ethanol for 30 mins each time. Subsequently, the samples were washed by dipping into ethanol multiple times to remove the by-products and unreacted residues, followed by air drying with nitrogen. To further enhance the electrical conductivity of the samples, we used acid treatment by soaking the samples in 0.5 M $H_2SO_4$ for 5 min at room temperature followed by heating the samples at 150 °C for another 10 min. On acid treatment, the OTf counterions in the PEDOT:OTf films were replaced by sulfate counterions ($HSO_4^-$), which was confirmed from the removal of fluorine signals in the XPS results (Supplementary Fig. 17).

*Nanoantenna fabrication*
A modified version of colloidal lithography was used for the nanodisk array fabrication as shown in Supplementary Fig. 4[56]. To summarize the process, a 4 wt% PMMA ($M_w \approx 996,000$ (Sigma-Aldrich)) solution in anisole (Sigma Aldrich) was spin coated at 5000 rpm onto the PEDOT:Sulf thin films. Soft baking at 140 °C for 10 min was then applied. The samples were treated with reactive oxygen plasma (50 W, 250 mtorr) for 20 s to increase the hydrophilicity of the surface. To functionalize the PMMA surface to be positively charged, 2 wt% poly(diallyldimethylammonium chloride) (PDDA) (522376 (Sigma-Aldrich)) in deionized water was drop-cast onto the film to fully cover it. After 1 min 30 s, the samples were rinsed with deionized water for 40 s, followed by air drying with nitrogen. Negatively charged polystyrene nanoparticles (PS beads, 197 nm, 0.2–0.3 wt% in deionized water (Microparticles GmbH, PS-ST KM56-1)) were then dropped on the samples. After 10 min, the samples coated with the PS beads were rinsed with deionized water and dried with a nitrogen stream, which produced a sparse monolayer of PS beads on the PMMA/PEDOT:Sulf thin films. This was followed by reactive oxygen plasma etching (250 mtorr, 50 W) for 50 - 130 s, using the PS beads monolayer as the mask. Depending on the thicknesses of the PMMA and PEDOT:Sulf thin films, the time interval of etching was varied to ensure a complete removal of the PMMA and PEDOT:Sulf parts that were not covered by the mask. The samples were then placed into an acetone bath and soaked for 30 min followed by a mild sonication for 5 min and nitrogen air drying to remove the PMMA and PS beads to finally produce the PEDOT:Sulf nanodisks.

*Electrochemically tunable nanoantenna fabrication*
The ion gel was prepared by adapting the procedure described in literature[42]. It entailed making a solution mixture of the co-polymer poly(vinylidene fluoride-co-hexafluoropropylene) (PVDF-HFP, $M_n = 130,000$, Sigma Aldrich) in acetone at a 1:7 weight ratio. The solution mixture was stirred overnight at 50 °C to ensure complete dissolution of the PVDF-HFP in acetone. The ionic liquid 1-ethyl-3-methylimidazolium bis(trifluoro-methylsulfonyl)imide ([EMIM$^+$][TFSI$^-$], Sigma Aldrich) was then added into the PVDF-HFP:Acetone mixture at a 1:2 ratio and the mixture was stirred on a hotplate at 50 °C for 30 min. The ion gel mixture was spin-coated onto the glass/ITO substrate with PEDOT:Sulf nanodisks at 1000 rpm for 1 min. The spin-coated ion gel was then directly placed on a hotplate at 60 °C for 2 hours to ensure drying. Another precleaned glass/ITO substrate was then placed on top of the ion gel, with the ITO contacting the ion gel, in order to function as the top electrode used to apply biases to the ion gel. The two ITO substrates were then taped together towards the edges and clamped with crocodile clips to ensure good contact between the top ITO electrode and the ion gel before placing them in the UV-Vis-NIR substrate holder.

*UV-Vi-NIR spectroscopy for measuring voltage dependent extinction spectra, switching, and chronocoulometry*
The extinction spectra in the vis–NIR range (400–3300 nm) were measured using a UV–vis–NIR spectrometer (Lambda 900 (Perkin Elmer Instruments)). The extinction spectra include transmission losses due to both absorption and scattering. For static voltage dependent measurements, a specific voltage was applied for 3-5 min to ensure the stabilization of the current before recording the spectra. A blank spectra for the voltage dependent measurements was a sample containing everything besides the PEDOT:Sulf nanodisks, specifically: glass/ITO/ion gel/ITO/glass. For samples with PEDOT:Sulf films or nanodisks on glass/ ITO substrates, glass/ITO was used as the reference. For the switching measurements, time-drive mode was used to record changes in extinction over time upon switching. The bottom ITO contact was grounded, and the top ITO contact was used to apply potentials to the substrates with a Kiethley 2400. To measure the current over time with various biases, an electrochemical potentiostat was used to apply and record the changes in current over a time period of 3 minutes until the current stabilized.

*Ellipsometry*
PEI vapour-reduced PEDOT:Sulf film deposited on a 2-inch single-side polished *c*-plane sapphire wafer was characterized under normal ambient conditions at room temperature. Spectroscopic ellipsometry data were collected by ellipsometers with different spectral ranges at two to four incident angles (UV-Vis-NIR: 0.8 eV to 5.9 eV at 40°, 50°, 60°, and 70°; MIR: 0.04 eV to 0.80 eV at 40° and 60°; and THz: 0.0028 eV to 0.0040 eV at 40°, 50°, and 60°). The detailed information of the characterization can be found in our previous study[30,52]. The obtained data were analyzed by WVASE software (*J. A. Woollam Co.*) and an anisotropic Drude-Lorentz model was utilized for the reduced PEDOT:Sulf film with its thickness determined by surface profiler. The extracted permittivity and corresponding fitting parameters for the reduced PEDOT:Sulf film is shown in the Supplementary Fig. 1 and Supplementary Table 1.

*Electrical, chemical, and morphological characterization*
Sheet resistance, $R_s$, of thin films were measured using a four-point probe set-up with a Signatone Pro4 S-302 resistivity stand and a Keithley 2400. The film thickness $t$ was determined by Veeco Dimension 3100 AFM. This was used to measure the electrical conductivity of PEDOT:OTf and PEDOT:Sulf films as reported in Supplementary Table 3. Surface morphology of the nanodisks was obtained by Veeco Dimension 3100 AFM and the images were analyzed using Nanoscope Analysis software (Bruker). X-ray photoemission experiments were carried out using a Scienta ESCA 200 spectrometer under ultrahigh vacuum conditions at a base pressure of $1 \times 10^{-10}$ mbar. The XPS measurements were done with a monochromatic Al Kα X-ray source, which provided photons with an energy of 1,486.6 eV.

*Optical numerical simulations and calculations*
Numerical simulations of the electromagnetic response of PEDOT:Sulf nanoantennas were performed *via* the finite-difference time-domain (FDTD) method using the commercial software Lumerical FDTD Solutions (http://www.lumerical.com/fdtd.php). The optical parameters for the oxidized and chemically reduced PEDOT:Sulf thin films were taken as the anisotropic complex permittivity obtained from the ellipsometry measurements. For periodic nanodisk arrays and thin films, the spectra and near-field profiles were recorded via field and power monitors. Periodic PEDOT:Sulf nanodisk arrays (or thin film) were placed on top of the glass substrates. The structures were illuminated by a plane wave light source at normal incidence. Antisymmetrical and symmetrical boundaries were used for the *x* axis (parallel to the polarization) and *y* axis (normal to the polarization) and perfectly matched layers were used for the *z* axis (parallel to the light incident direction). For single nanodisks, the spectra were obtained using a total field/scattered field method and by extracting the extinction cross-section of isolated PEDOT:Sulf nanodisks on a glass substrate. Geometry parameters are indicated in each graph (diameter, thickness and/or array period) and the mesh size was $1 \times 1 \times 1$ nm$^3$. The optical parameter for glass[30] were taken from the literature and permittivities of PEDOT:Sulf was determined by ellipsometry. In the analytical calculations, the effective permittivity of the surroundings was calculated based on an average refractive index of ion gel and glass ($\varepsilon_s = [(n_{iongel} + n_{glass})/2]^2$).

## 5. Acknowledgements

The authors acknowledge financial support from the Knut and Alice Wallenberg Foundation, the Swedish Research Council (VR), the Swedish Foundation for Strategic Research (SSF), and the Swedish Government Strategic Research Area in Materials Science on Functional Materials at Linköping University (Faculty Grant SFO-Mat-LiU No 2009 00971). M.P.J. is a Wallenberg Academy Fellow. The authors would like to thank Dr. Dan Zhao and Dr. Ludovico Migliaccio for helpful discussions.

**TOC image**

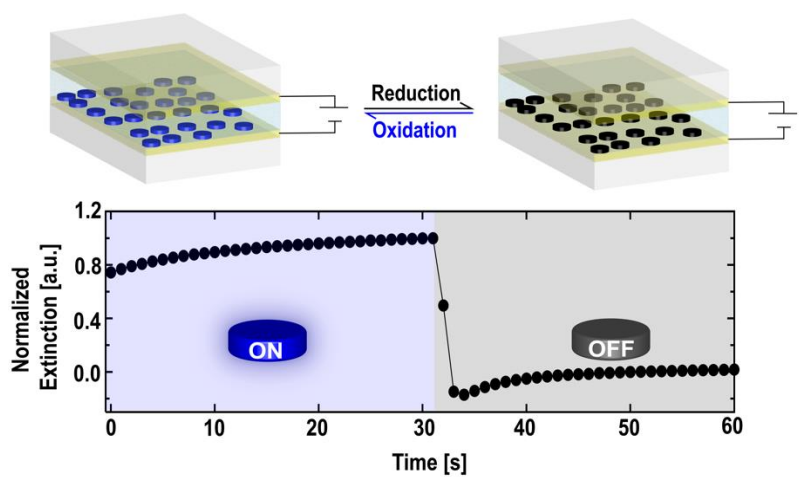

## 6. References

1. Novotny, L. & van Hulst, N. Antennas for light. *Nat. Photonics* **5**, 83–90 (2011).
2. Shi, X. *et al.* Enhanced water splitting under modal strong coupling conditions. *Nat. Nanotechnol.* **13**, 953–958 (2018).
3. Atwater, H. A. & Polman, A. Plasmonics for improved photovoltaic devices. *Nat. Mater.* **9**, 205–213 (2010).
4. Sheldon, M. T., Groep, J. van de, Brown, A. M., Polman, A. & Atwater, H. A. Plasmoelectric potentials in metal nanostructures. *Science* **346**, 828–831 (2014).
5. Yesilkoy, F. *et al.* Phase-sensitive plasmonic biosensor using a portable and large field-of-view interferometric microarray imager. *Light Sci. Appl.* **7**, 17152–17152 (2018).
6. Mazzotta, F., Wang, G., Hägglund, C., Höök, F. & Jonsson, M. P. Nanoplasmonic biosensing with on-chip electrical detection. *Biosens. Bioelectron.* **26**, 1131–1136 (2010).
7. Jonsson, M. P., Dahlin, A. B., Jönsson, P. & Höök, F. Nanoplasmonic biosensing with focus on short-range ordered nanoholes in thin metal films (Review). *Biointerphases* **3**, FD30–FD40 (2008).
8. Jonsson, M. P., Jönsson, P., Dahlin, A. B. & Höök, F. Supported Lipid Bilayer Formation and Lipid-Membrane-Mediated Biorecognition Reactions Studied with a New Nanoplasmonic Sensor Template. *Nano Lett.* **7**, 3462–3468 (2007).
9. Englebienne, P. Use of colloidal gold surface plasmon resonance peak shift to infer affinity constants from the interactions between protein antigens and antibodies specific for single or multiple epitopes. *Analyst* **123**, 1599–1603 (1998).
10. Shao, L., Zhuo, X. & Wang, J. Advanced Plasmonic Materials for Dynamic Color Display. *Adv. Mater.* **30**, 1704338 (2018).

11. Franklin, D., Frank, R., Wu, S.-T. & Chanda, D. Actively addressed single pixel full-colour plasmonic display. *Nat. Commun.* **8**, 15209 (2017).
12. Xiong, K. *et al.* Plasmonic Metasurfaces with Conjugated Polymers for Flexible Electronic Paper in Color. *Adv. Mater.* **28**, 9956–9960 (2016).
13. Xiong, K., Tordera, D., Jonsson, M. P. & Dahlin, A. B. Active control of plasmonic colors: emerging display technologies. *Rep. Prog. Phys.* **82**, 024501 (2019).
14. Park, J., Kang, J.-H., Kim, S. J., Liu, X. & Brongersma, M. L. Dynamic Reflection Phase and Polarization Control in Metasurfaces. *Nano Lett.* **17**, 407–413 (2017).
15. Chen, X. *et al.* Dual-polarity plasmonic metalens for visible light. *Nat. Commun.* **3**, 1198 (2012).
16. Aieta, F. *et al.* Aberration-Free Ultrathin Flat Lenses and Axicons at Telecom Wavelengths Based on Plasmonic Metasurfaces. *Nano Lett.* **12**, 4932–4936 (2012).
17. Pendry, J. B. Negative Refraction Makes a Perfect Lens. *Phys. Rev. Lett.* **85**, 3966–3969 (2000).
18. Yu, N. *et al.* Light Propagation with Phase Discontinuities: Generalized Laws of Reflection and Refraction. *Science* **334**, 333–337 (2011).
19. Naik, G. V., Shalaev, V. M. & Boltasseva, A. Alternative Plasmonic Materials: Beyond Gold and Silver. *Adv. Mater.* **25**, 3264–3294 (2013).
20. Wang, Y. *et al.* Electrical tuning of phase-change antennas and metasurfaces. *Nat. Nanotechnol.* 1–6 (2021) doi:10.1038/s41565-021-00882-8.
21. Alaee, R., Albooyeh, M., Tretyakov, S. & Rockstuhl, C. Phase-change material-based nanoantennas with tunable radiation patterns. *Opt. Lett.* **41**, 4099–4102 (2016).
22. Wuttig, M., Bhaskaran, H. & Taubner, T. Phase-change materials for non-volatile photonic applications. *Nat. Photonics* **11**, 465–476 (2017).
23. Kim, Y. *et al.* Phase Modulation with Electrically Tunable Vanadium Dioxide Phase-Change Metasurfaces. *Nano Lett.* **19**, 3961–3968 (2019).
24. Abdollahramezani, S. *et al.* Tunable nanophotonics enabled by chalcogenide phase-change materials. *Nanophotonics* **9**, 1189–1241 (2020).
25. Liu, Z. *et al.* Tuning infrared plasmon resonances in doped metal-oxide nanocrystals through cation-exchange reactions. *Nat. Commun.* **10**, 1394 (2019).
26. Fang, Z. *et al.* Gated Tunability and Hybridization of Localized Plasmons in Nanostructured Graphene. *ACS Nano* **7**, 2388–2395 (2013).
27. Yan, H. *et al.* Tunable infrared plasmonic devices using graphene/insulator stacks. *Nat. Nanotechnol.* **7**, 330–334 (2012).
28. Ju, L. *et al.* Graphene plasmonics for tunable terahertz metamaterials. *Nat. Nanotechnol.* **6**, 630–634 (2011).
29. Reynolds, J. R., Thompson, B. C. & Skotheim, T. A. *Conjugated Polymers: Properties, Processing, and Applications*. (CRC Press, 2019). doi:10.1201/9780429190520.
30. Chen, S. *et al.* Conductive polymer nanoantennas for dynamic organic plasmonics. *Nat. Nanotechnol.* **15**, 35–40 (2020).
31. Bohn, B. J. *et al.* Near-Field Imaging of Phased Array Metasurfaces. *Nano Lett.* **15**, 3851–3858 (2015).
32. Ke, Y. *et al.* Adaptive Thermochromic Windows from Active Plasmonic Elastomers. *Joule* **3**, 858–871 (2019).
33. Zozoulenko, I. *et al.* Polarons, Bipolarons, And Absorption Spectroscopy of PEDOT. *ACS Appl. Polym. Mater.* **1**, 83–94 (2019).
34. Sonmez, G. Polymeric electrochromics. *Chem. Commun.* 5251–5259 (2005) doi:10.1039/B510230H.
35. Ibanez, J. G. *et al.* Conducting Polymers in the Fields of Energy, Environmental Remediation, and Chemical–Chiral Sensors. *Chem. Rev.* **118**, 4731–4816 (2018).

36. Massonnet, N., Carella, A., Geyer, A. de, Faure-Vincent, J. & Simonato, J.-P. Metallic behaviour of acid doped highly conductive polymers. *Chem. Sci.* **6**, 412–417 (2014).
37. Rivnay, J. *et al.* Organic electrochemical transistors. *Nat. Rev. Mater.* **3**, 1–14 (2018).
38. Kim, D., Jang, H., Lee, S., Kim, B. J. & Kim, F. S. Solid-State Organic Electrolyte-Gated Transistors Based on Doping-Controlled Polymer Composites with a Confined Two-Dimensional Channel in Dry Conditions. *ACS Appl. Mater. Interfaces* **13**, 1065–1075 (2021).
39. Hasse, B. *et al.* Viscosity, Interfacial Tension, Density, and Refractive Index of Ionic Liquids [EMIM][MeSO3], [EMIM][MeOHPO2], [EMIM][OcSO4], and [BBIM][NTf2] in Dependence on Temperature at Atmospheric Pressure. *J. Chem. Eng. Data* **54**, 2576–2583 (2009).
40. Porter, J. M. *et al.* Optical measurements of impurities in room-temperature ionic liquids. *J. Quant. Spectrosc. Radiat. Transf.* **133**, 300–310 (2014).
41. Desai, S. *et al.* Gel electrolytes with ionic liquid plasticiser for electrochromic devices. *Electrochimica Acta* **56**, 4408–4413 (2011).
42. Zhao, D. *et al.* Polymer gels with tunable ionic Seebeck coefficient for ultra-sensitive printed thermopiles. *Nat. Commun.* **10**, 1093 (2019).
43. Cendra, C. *et al.* Role of the Anion on the Transport and Structure of Organic Mixed Conductors. *Adv. Funct. Mater.* **29**, 1807034 (2019).
44. Xiong, K., Olsson, O., Svirelis, J., Baumberg, J. & Dahlin, A. Video Speed Switching of Plasmonic Structural Colors with High Contrast and Superior Lifetime. *Adv. Mater. Accept.*
45. Brooke, R. *et al.* Organic energy devices from ionic liquids and conducting polymers. *J. Mater. Chem. C* **4**, 1550–1556 (2016).
46. Chen, S. *et al.* Tunable Structural Color Images by UV-Patterned Conducting Polymer Nanofilms on Metal Surfaces. *Adv. Mater.* **n/a**, 2102451.
47. Kawahara, J., Ersman, P. A., Engquist, I. & Berggren, M. Improving the color switch contrast in PEDOT:PSS-based electrochromic displays. *Org. Electron.* **13**, 469–474 (2012).
48. Tehrani, P., Hennerdal, L.-O., Dyer, A. L., Reynolds, J. R. & Berggren, M. Improving the contrast of all-printed electrochromic polymer on paper displays. *J. Mater. Chem.* **19**, 1799–1802 (2009).
49. Maier, S. A. *Plasmonics: Fundamentals and Applications*. (Springer Science & Business Media, 2007).
50. Wokaun, A., Gordon, J. P. & Liao, P. F. Radiation Damping in Surface-Enhanced Raman Scattering. *Phys. Rev. Lett.* **48**, 957–960 (1982).
51. Langhammer, C., Yuan, Z., Zorić, I. & Kasemo, B. Plasmonic Properties of Supported Pt and Pd Nanostructures. *Nano Lett.* **6**, 833–838 (2006).
52. Chen, S. *et al.* On the anomalous optical conductivity dispersion of electrically conducting polymers: ultra-wide spectral range ellipsometry combined with a Drude–Lorentz model. *J. Mater. Chem. C* **7**, 4350–4362 (2019).
53. Paulsen, B. D. *et al.* Time-Resolved Structural Kinetics of an Organic Mixed Ionic–Electronic Conductor. *Adv. Mater.* **32**, 2003404 (2020).
54. Thomas, E. M. *et al.* X-Ray Scattering Reveals Ion-Induced Microstructural Changes During Electrochemical Gating of Poly(3-Hexylthiophene). *Adv. Funct. Mater.* **28**, 1803687 (2018).
55. Brooke, R. *et al.* Vapor phase synthesized poly(3,4-ethylenedioxythiophene)-trifluoromethanesulfonate as a transparent conductor material. *J. Mater. Chem. A* **6**, 21304–21312 (2018).

56. Hanarp, P., Käll, M. & Sutherland, D. S. Optical Properties of Short Range Ordered Arrays of Nanometer Gold Disks Prepared by Colloidal Lithography. *J. Phys. Chem. B* **107**, 5768–5772 (2003).

# Supplementary Information for

## Electrical Tuning of Plasmonic Conducting Polymer Nanoantennas

Akchheta Karki[1], Giancarlo Cincotti[1], Shangzhi Chen[1], Chuanfei Wang[1], Vallery Stanishev[2,3], Vanya Darakchieva[2,3], Mats Fahlman[1], Magnus P. Jonsson[1]★

[1]Laboratory of Organic Electronics, Department of Science and Technology (ITN), Linköping University, SE-601 74 Sweden
[2]Terahertz Materials Analysis Center (THeMAC), Department of Physics, Chemistry and Biology (IFM), Linköping University, Linköping, SE-581 83 Sweden
[3]Center for III-Nitride Technology, C3NiT-Janzèn, Department of Physics, Chemistry and Biology (IFM), Linköping University, Linköping, SE-581 83 Sweden

★**Corresponsence: magnus.jonsson@liu.se**

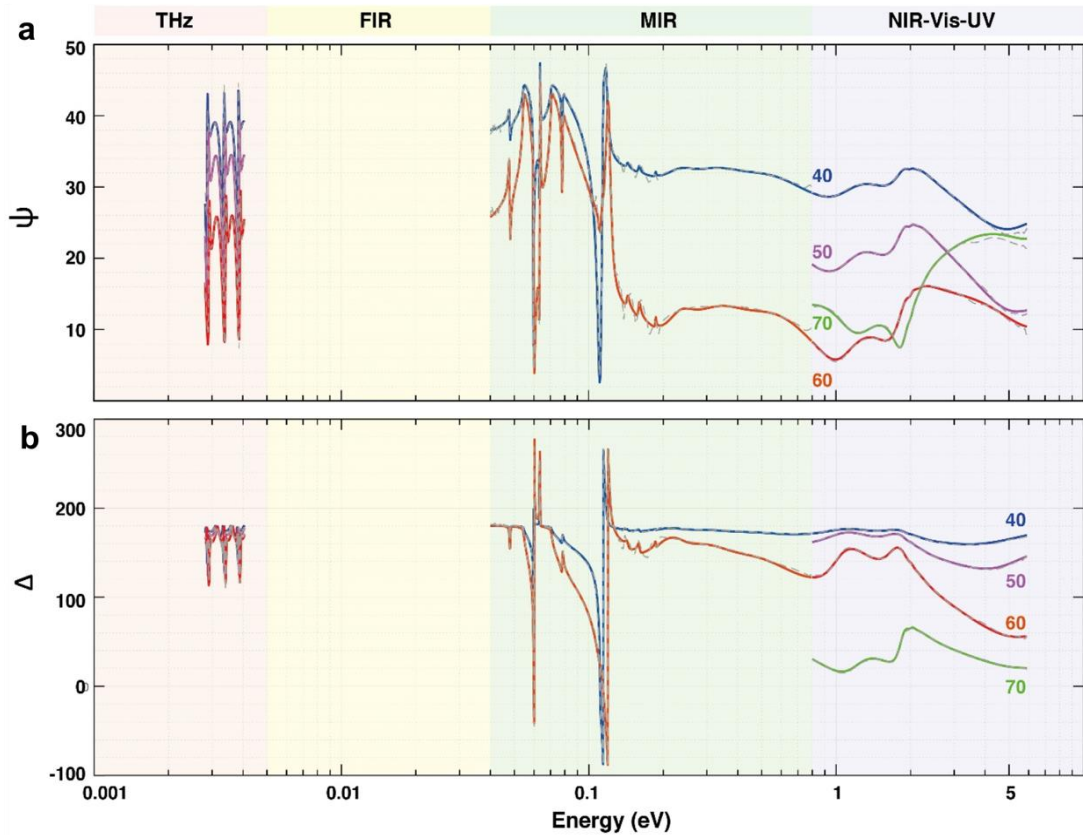

**Supplementary Fig. 1 | Spectroscopic ellipsometry for PEI vapor-reduced PEDOT:Sulf film.** Ellipsometric raw data ψ **a**, and Δ **b**, of PEI vapor-reduced PEDOT:Sulf film in the spectral range from 2.8 meV (0.67 THz) to 5.90 eV (210 nm), including THz, MIR, and NIR-Vis-UV three regions. The experimental data are shown in solid lines and the best-matched calculated data using the anisotropic Drude-Lorentz model[1] are in dashed lines. The best-matched parameters obtained from the ellipsometry data analysis are listed in Supplementary Table 1.

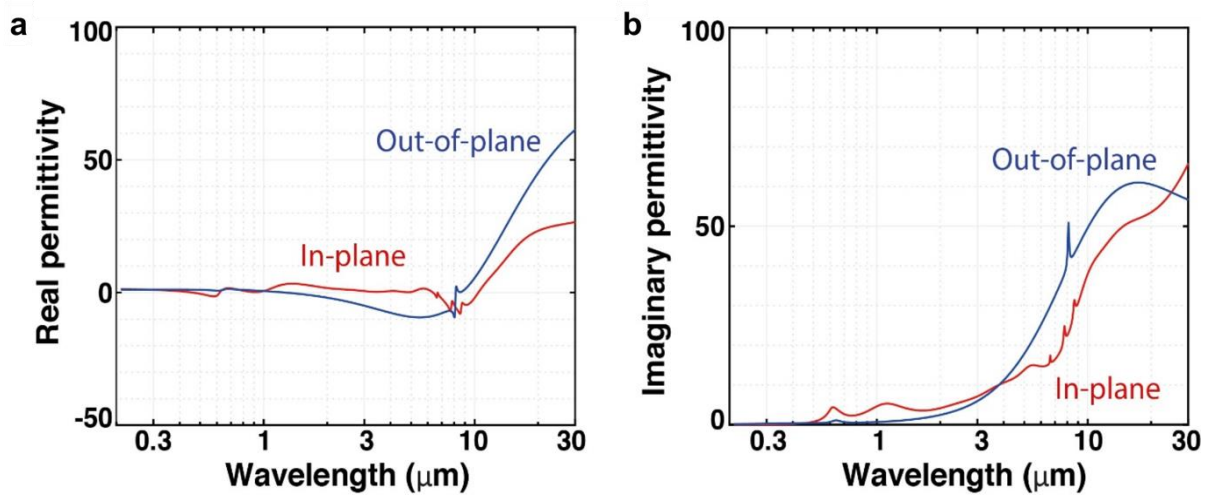

**Supplementary Fig. 2 | Permittivity dispersion for PEI vapor-reduced PEDOT:Sulf film. a,** Real permittivity and **b,** imaginary permittivity dispersion curves obtained for PEI vapor-reduced PEDOT:Sulf film. Red curves are for the in-plane direction (x-y plane) and blue curves for the out-of-plane direction (z-axis). The extracted data are based on the Drude-Lorentz model

in Supplementary Fig. 1 and using the best-matched parameters presented in Supplementary Table 1.

**Supplementary Table 1 | Measured electrical properties of PEI vapor-reduced PEDOT:Sulf thin film.** Best-matched parameters obtained from the ellipsometry data analysis utilizing the anisotropic Drude-Lorentz model for PEI vapor-reduced PEDOT:Sulf film. Owing to the low sensitivity of the equipment for the poor electric conductivity sample, the Drude component of the model shows almost negligible contribution to the permittivity (ultra-large broadening and ultra-small amplitude). This is also obvious from the ellipsometric raw data in Supplementary Fig. 1, where in the THz spectral range, the spectra are almost identical with those of the sapphire substrate.

|  | Oscillator No. ($j^{th}$) | Frequency ω (eV) | Broadening γ (eV) | Amplitude A (eV$^2$) |
|---|---|---|---|---|
| | | $\varepsilon_\infty$ = 1.5462 | | |
| In-plane | 1 | 2.0222 | 0.1522 | 0.2583 |
| | 2 | 1.9854 | 0.4269 | 2.6275 |
| | 3 | 1.1589 | 0.6397 | 3.4621 |
| | 4 | 0.5275 | 0.4242 | 0.5863 |
| | 5 | 0.3259 | 0.2293 | 0.4058 |
| | 6 | 0.2331 | 0.0761 | 0.8266 |
| | 7 | 0.1865 | 0.0018 | 0.0008 |
| | 8 | 0.1599 | 0.0035 | 0.0026 |
| | 9 | 0.1435 | 0.0039 | 0.0024 |
| | 10 | 0.1258 | 0.0575 | 0.0869 |
| | 11 | 0.0947 | 0.0877 | 0.1817 |
| | 12 | 0.0198 | 0.1417 | 0.3318 |
| | Drude | 0.0000 | 7.2 x 10$^8$ | 0.0005 |
| | | $\varepsilon_\infty$ = 1.3052 | | |
| Out-of-plane | 1 | 3.9409 | 4.9005 | 5.0212 |
| | 2 | 1.9375 | 0.2379 | 0.3273 |
| | 3 | 0.1529 | 0.0028 | 0.0048 |
| | 4 | 0.1344 | 0.2547 | 1.3860 |
| | 5 | 0.0019 | 1.3869 | 1.0617 |
| | Drude | 0.0000 | 7.2 x 10$^8$ | 0.0005 |

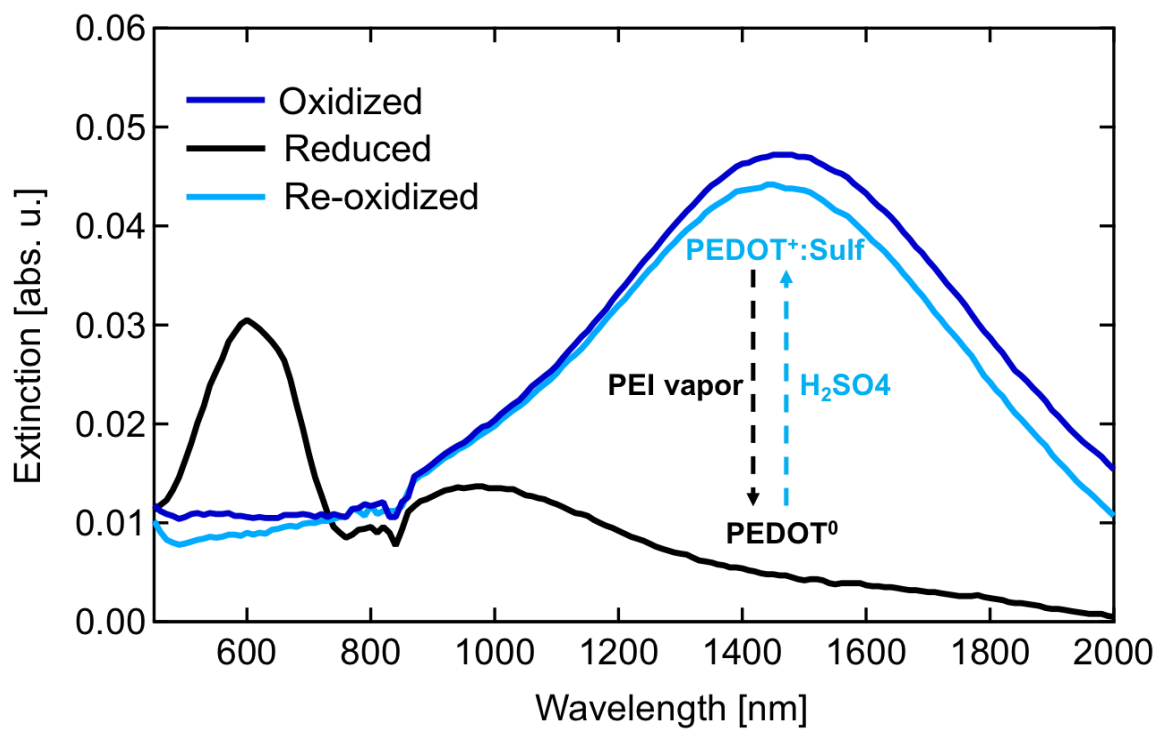

**Supplementary Fig. 3 | Experimental extinction of chemical and vapor treated PEDOT:Sulf at different redox states.** PEI vapor treatment was used to reduce the films and $H_2SO_4$ treatment was used to re-oxidize the film.

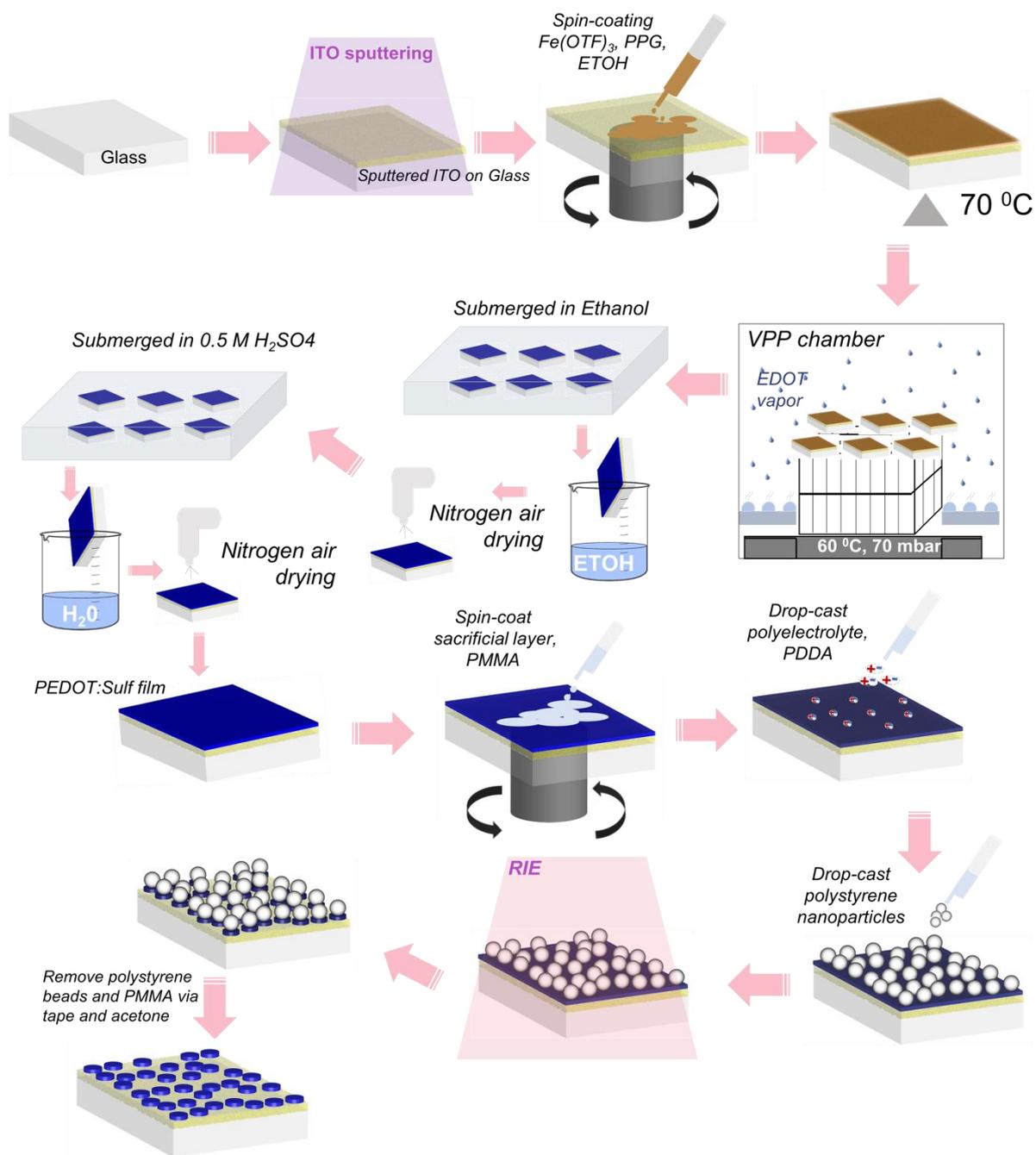

**Supplementary Fig. 4 | Fabrication procedure of PEDOT:Sulf nanodisks on glass/ITO substrates.** Detailed schematic showing the process to make nanodisks on glass/ITO substrates including vapor phase polymerization of PEDOT:Sulf polymer followed by colloidal lithography to make the PEDOT:Sulf nanostructures.

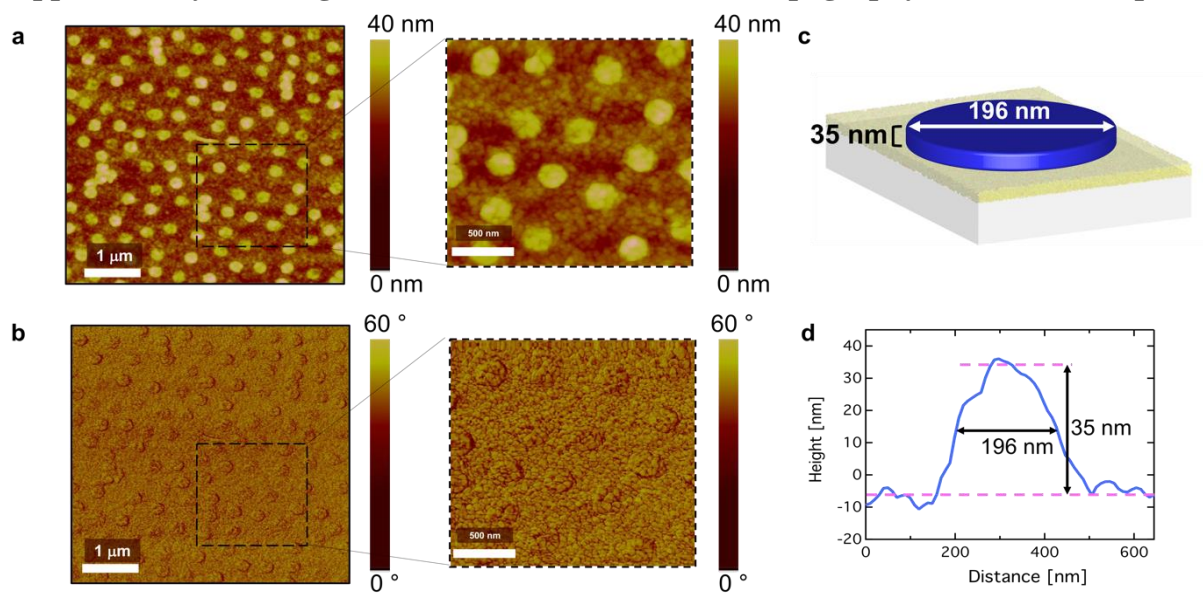

**Supplementary Fig. 5 | AFM topography and phase images for nanodisks made via 1× VPP.** 5×5 μm AFM height **a**, and phase **b**, images of 35 nm thick and 196 nm diameter PEDOT:Sulf nanodisks on glass/ITO substrates. Zoomed in images on the right are sized 2×2 μm. **c**, Schematic of height and width of a single nanodisk obtained from the height and width section analyses shown in **d**.

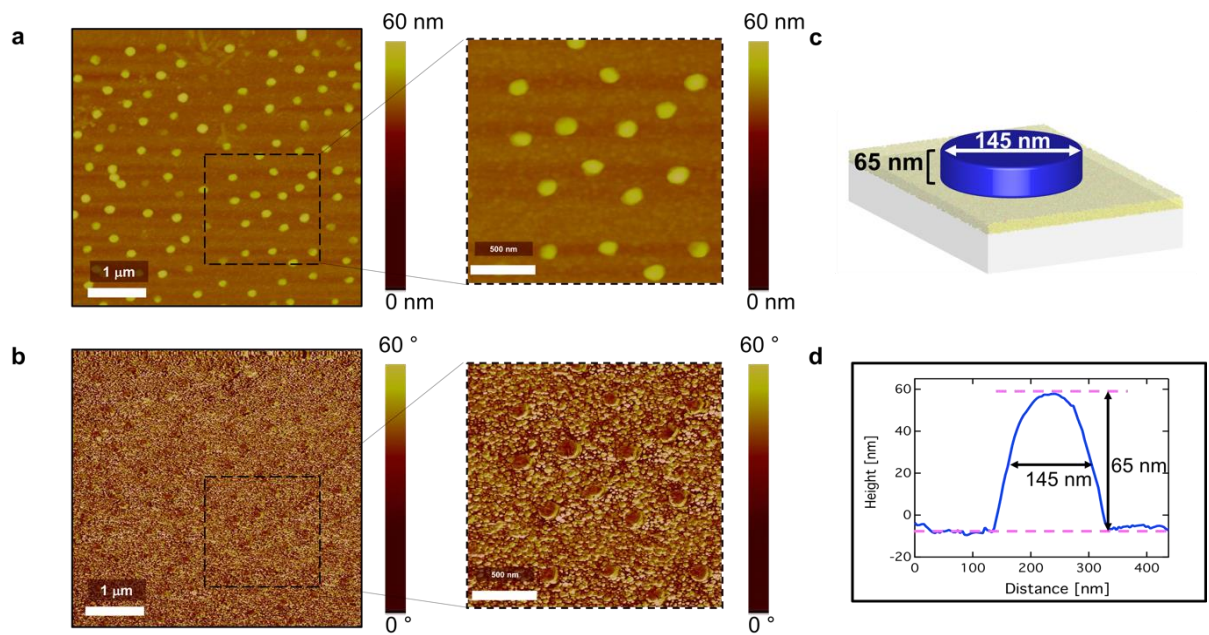

**Supplementary Fig. 6 | AFM topography and phase images for nanodisks made via 2× VPP.** 5×5 μm AFM height **a**, and phase **b**, images of 65 nm thick and 145 nm diameter PEDOT:Sulf nanodisks on glass/ITO substrates. Zoomed-in images on the right are sized 2×2 μm. **c**, Schematic of height and width of a single nanodisk obtained from the height and width section analyses shown in **d**.

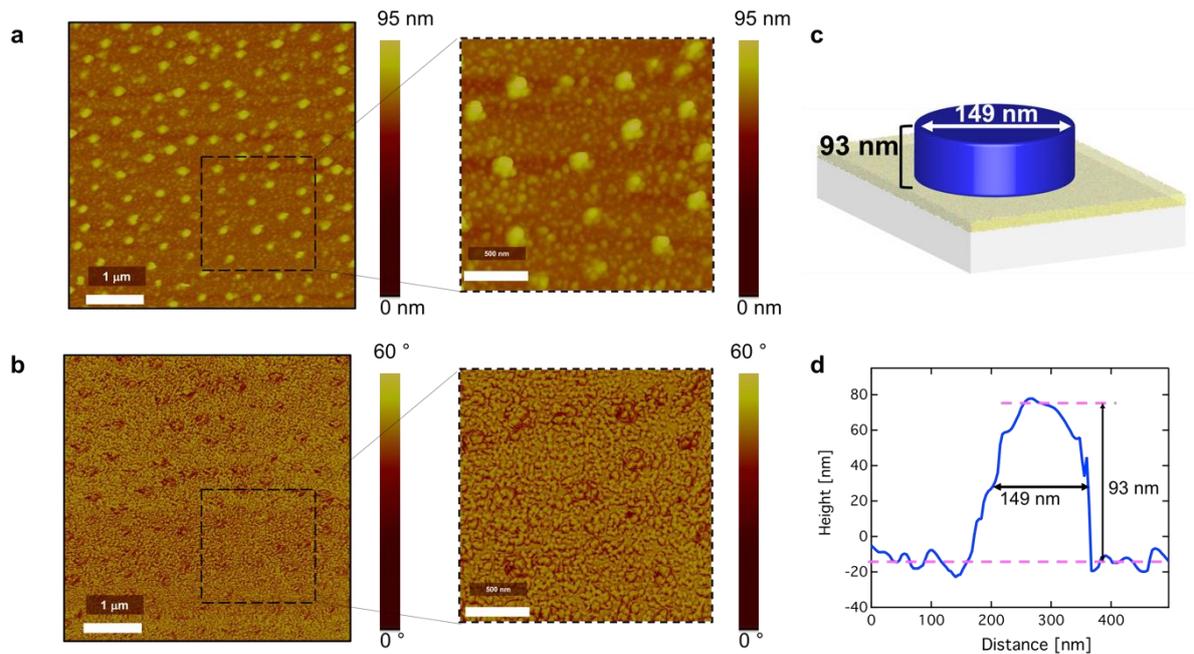

**Supplementary Fig. 7 | AFM topography and phase images for nanodisks made via 1× VPP.** 5×5 μm AFM height **a**, and phase **b**, images of 93 nm thick and 149 nm diameter PEDOT:Sulf nanodisks on glass/ITO substrates. Zoomed in images on the right are sized 2×2 μm. **c**, Schematic of height and width of a single nanodisk obtained from the height and width section analyses shown in **d**.

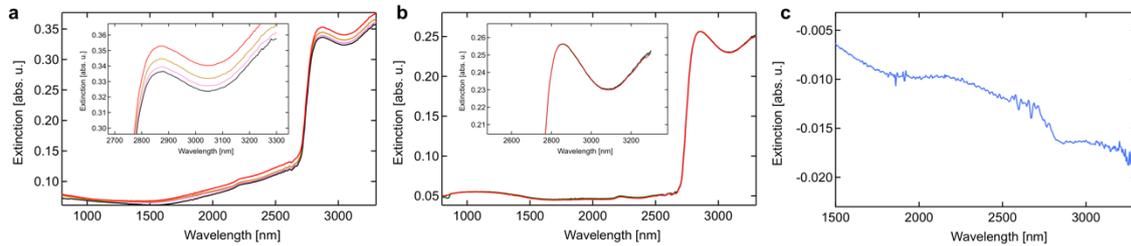

**Supplementary Fig. 8 | Extinction spectra of glass and ITO reference substrates. a**, Experimental extinction spectra of several different sputtered glass/ITO substrates also measured at different spots within the same substrate, which show variations in the extinction at higher wavelengths above 2000 nm. This is due to an unavoidable minor thickness variation that occurs during the ITO sputtering process (the inset shows zoomed-in spectra to highlight the differences in extinction). **b**, Experimental extinction spectra of a few glass substrates, where all spectra are essentially identical and overlap even when measured between different samples and spots (the inset shows zoomed-in spectra showing that there are no differences). Therefore, as a result of the slightly varying extinction spectra of ITO substrates, there can sometimes be negative extinction values at higher wavelengths as shown **c**, when two different thickness ITO substrates are subtracted from each other. The negative values are more conspicuous when measuring spectra with nanodisks which already have quite low extinction signals compared to the background signal in the 2000-3000 nm range, as well as comparable order of magnitude differences in the extinction values (0.01-0.03 abs. u).

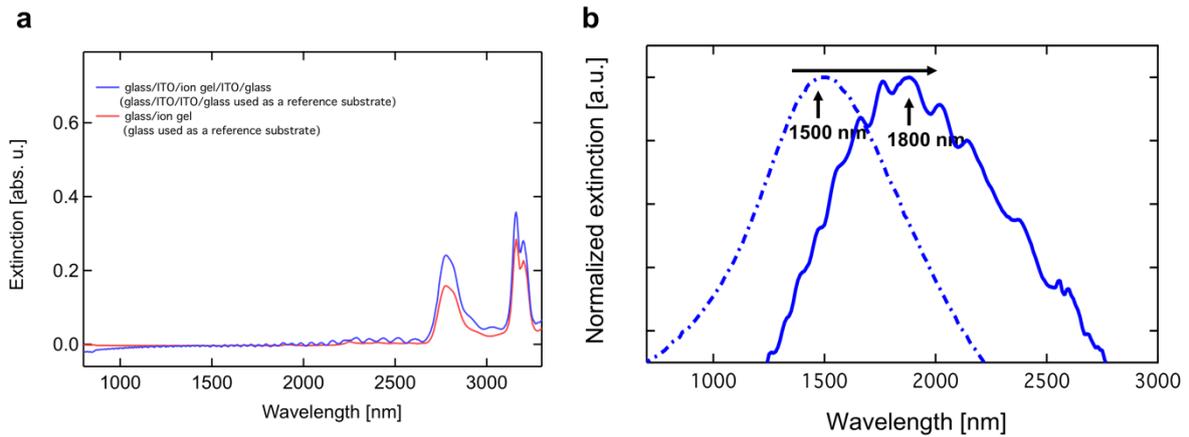

**Supplementary Fig. 9 | Extinction spectra of ion gel and its effect on the plasmonic resonance. a**, Experimental extinction spectra of ion gel on glass (red; glass is used as a reference) and in between two glass/ITO substrates (blue; same as used in the switching experiments, where glass/ITO/ITO/glass is used as a reference) showing that the undulations originate from the sandwiched ion gel structure and that they are more pronounced at increasing wavelengths. **b**, Experimental extinction spectra (normalized to the resonance peak intensity) of nanodisks (diameter=145 nm, thickness=65 nm) before (dashed blue line) and after spin-coating the ion gel and adding the ITO electrode on top, which shows that the resonance peak red-shifts as a result of the change in refractive index of the surrounding medium from that of air to that of the ion gel.

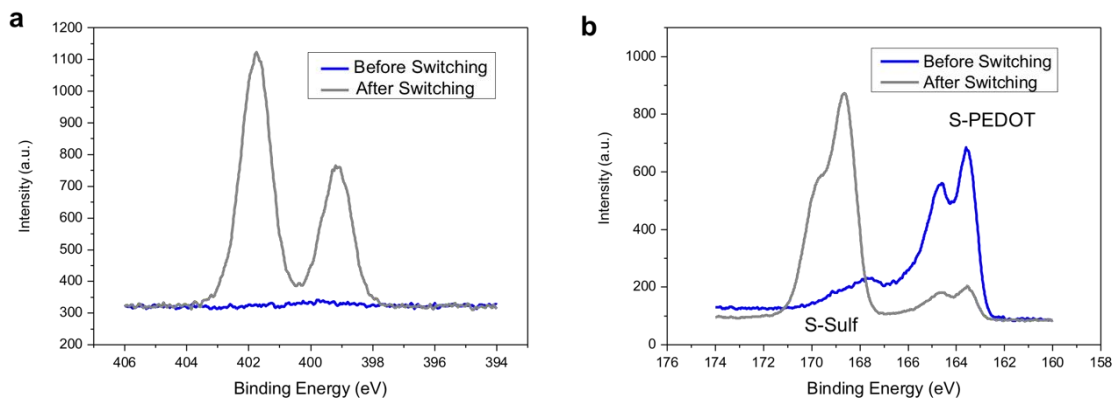

**Supplementary Fig. 10 | X-ray photoelectron spectroscopy for PEDOT thin films before and after electrical switching. a**, N1s X-ray photoelectron spectroscopy spectra for PEDOT:Sulf thin film before (blue curve) and after (grey curve) switching. **b,** S2p X-ray photoelectron spectroscopy spectra for PEDOT:Sulf thin film before (blue curve) and after (grey curve) switching.

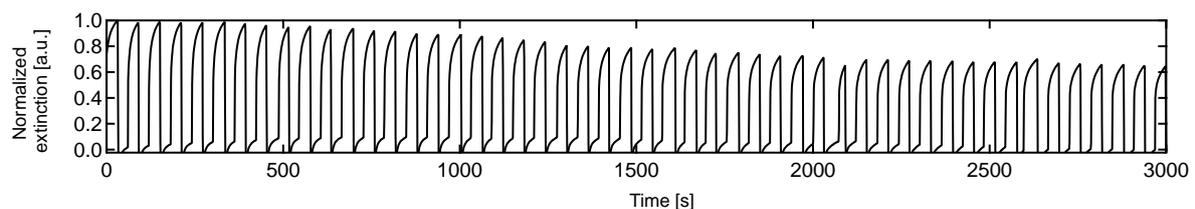

**Supplementary Fig. 11 | 50 minutes of on and off switching cycles.** Full 50 minutes of on and off cycles corresponding to the results in Fig. 3d in the main manuscript. The extinction

values are normalized to the on/off levels relative to the first switching cycle at the plasmonic peak position of ~1800 nm.

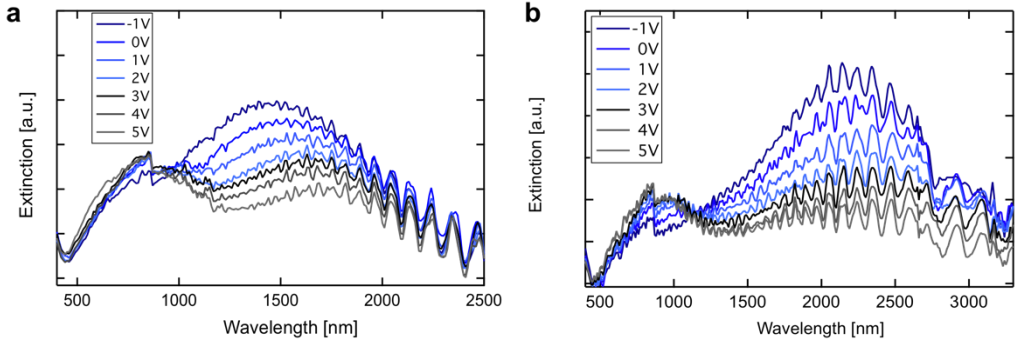

**Supplementary Fig. 12 | Bias dependent extinction spectra of PEDOT:Sulf nanodisks made with 1 × VPP and 3 × VPP.** Experimental extinction measurements at different biases for PEDOT:Sulf nanodisks made from 3 × VPP layers of PEDOT:Sulf (diameter=149 nm, thickness=93 nm, period=600 nm) **a,** and 1 × VPP layer of PEDOT:Sulf (diameter=196 nm, thickness=35 nm, period=600 nm) **b**. The undulations, which are more pronounced at longer wavelengths, are attributed to optical interference in the thin film device structure due to the sandwiched ion gel structure. This was confirmed from experimentally measured extinction spectra of equivalent devices without the nanodisks but with the ion gel between two ITO substrates (Supplementary Fig. 9a). Similar undulations have also been reported for similar ion gel components and device structures as used in our work[2].

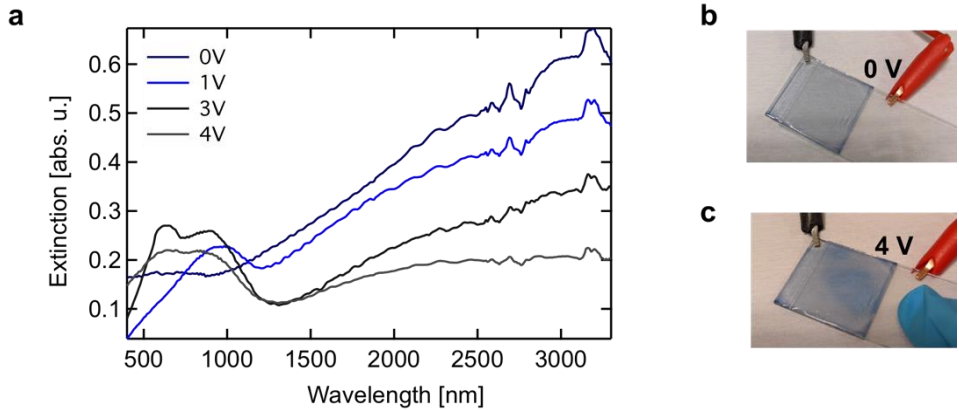

**Supplementary Fig. 13 | Electrical tuning of PEDOT:Sulf films. a**, Measured extinction spectra of PEDOT:Sulf films at different biases**,** photos of PEDOT:Sulf film with ion gel and ITO electrode on top at 0V (**b**), and 4V (**c**). The film turns a dark blue color with applied voltages indicating the reduction of PEDOT:Sulf to reduced PEDOT.

**Supplementary Note A:**

The free charge carrier density and DC mobility of the Drude model are related to the plasma frequency and momentum-averaged scattering time as[3]:

$$\omega_p = \sqrt{\frac{ne^2}{m\varepsilon_0}} \tag{S1}$$

$$\tau = \frac{m\mu}{e} \tag{S2}$$

where $n$ is the free charge carrier concentration, $e$ is the elemental charge, and $m$ is the effective mass.

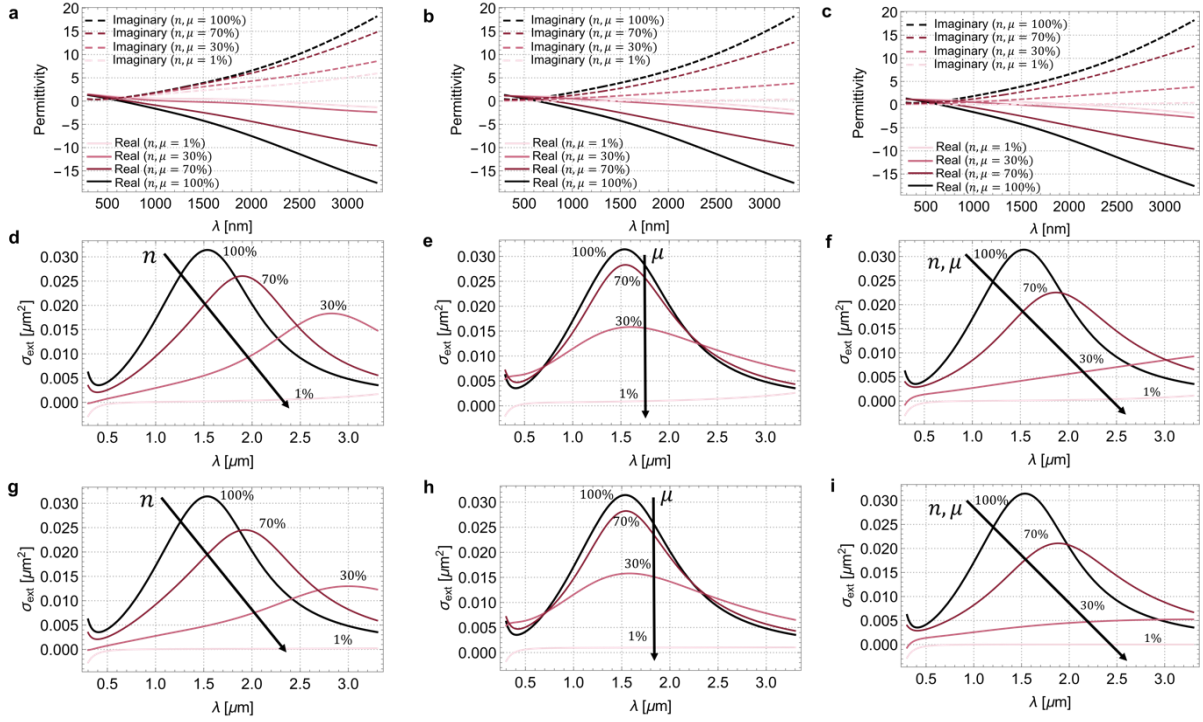

**Supplementary Fig. 14 | Calculations showing in-plane permittivity and extinction upon changing $n$, $\mu$, $A_j$, $\gamma_j$ in the Drude-Lorentz equation (Equation 1). a,** Real (solid lines) and imaginary (dashed lines) in-plane permittivity upon changing both $n$ and $\mu$ by 100%, 70%, 30%, and 1%, considering only the Drude-term in the Drude-Lorentz equation corresponding to the calculations and simulations shown in Fig. 4 in the main manuscript. (**b,c**), Real and imaginary in-plane permittivity if reducing both $n$, $\mu$, $A_j$, $\gamma_j^{-1}$ from 100% to 70%, 30%, and 1% considering **b**: the Drude term and the first three Lorentz oscillators (as previously reported part of the transport function and shown in Supplementary Table 1)[4] and **c**: all Lorentz oscillators. **d-f**, Calculated extinction based on dipolar polarizability theory of nanodisks treated as oblate spheroids, considering changes to the Drude term and the first three Lorentz oscillators upon: varying only $n$ and $A_j$ (d), varying only $\mu$ and $\gamma_j^{-1}$ (e), and (f) varying both $n$, $\mu$, $A_j$, $\gamma_j^{-1}$ (at 100%, 70%, 30%, and 1%). **g-i**, Same as in (d-f) but considering all Lorentz oscillators.

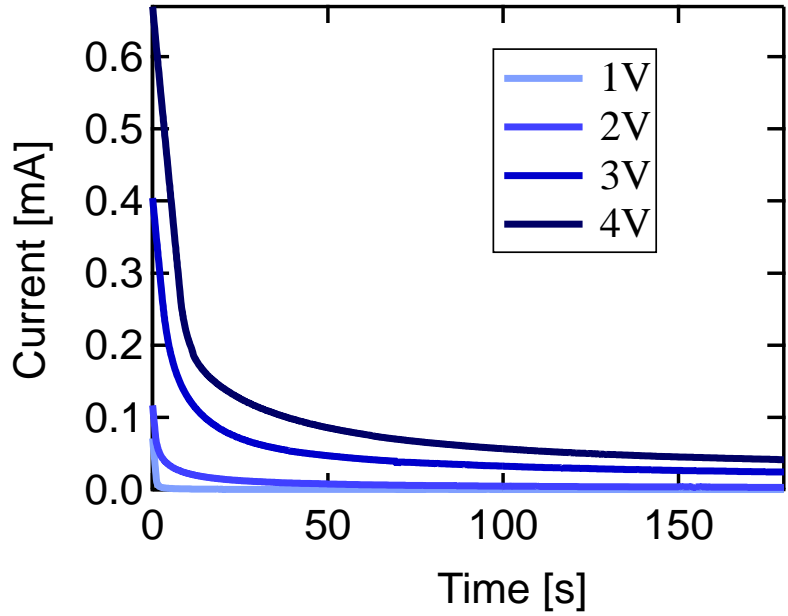

**Supplementary Fig. 15 | Measurement of mobile polaronic carrier density *via* chronocoulometry.** Measured current over time upon reducing PEDOT:Sulf films using to estimate the polaronic charge carrier density.

**Supplementary Table 2 | Estimation of mobile polaronic charge carrier density measured *via* chronocoulometry.** Estimation of mobile polaronic charge carrier density ($n$) by applying a bias and integrating the current over time [charge [Q]= current [I] × time [t]] and dividing by the volume [$n = Q/q*volume$], where volume = area of the film × thickness.
*The polaronic charge carrier density of the fully oxidized film was set to $2.6 \times 10^{21}$ cm$^{-3}$ as determined by ellipsometry and the Drude-Lorentz model[4].

| Bias [V] | Integrated $n$ from potentiostat [cm$^{-3}$] | % of $n$ relative to fully oxidized film* [%] | Remaining $n$ in film after applied bias [cm$^{-3}$] | % of remaining $n$ relative to fully oxidized film* |
|---|---|---|---|---|
| 1 | $2.25 \times 10^{19}$ | 0.87 | $2.57 \times 10^{21}$ | 99.1 |
| 2 | $2.40 \times 10^{20}$ | 9.20 | $2.35 \times 10^{21}$ | 90.8 |
| 3 | $1.22 \times 10^{21}$ | 46.8 | $1.21 \times 10^{21}$ | 53.2 |
| 4 | $2.32 \times 10^{21}$ | 89.1 | *$2.80 \times 10^{20}$ | 10.9 |

**Supplementary Note B:**

We can estimate the mobility of the reduced PEDOT:Sulf using $\sigma = qn\mu$. The electrical conductivity of a PEI-reduced film ($\sigma_{red}$) was determined as the average from 4-point probe measurements, $\sigma_{red} = 9.975 \pm 1.73$ S/cm. The carrier density for the reduced film was set to $n_{red} = 2.80 \times 10^{20}$ cm$^{-3}$ as obtained from Supplementary Table 1 and $q = 1.6 \times 10^{-19}$ C.

$$\mu_{red} = \frac{\sigma_{red}}{qn_{red}}$$

This gives an estimated $\mu_{red} = 0.199 \pm 0.035$ cm$^2$/Vs

The mobility for the oxidized film was taken from our previous report[4]: $\mu_{ox}$= 13.4 cm$^2$/Vs
This gives that the mobility of the reduced film is (0.199/13.4)*100 = 1.5% of the oxidized film. Notably, this value is likely underestimated since the PEI-reduced film may have lower carrier density than the electrochemically reduced materials.

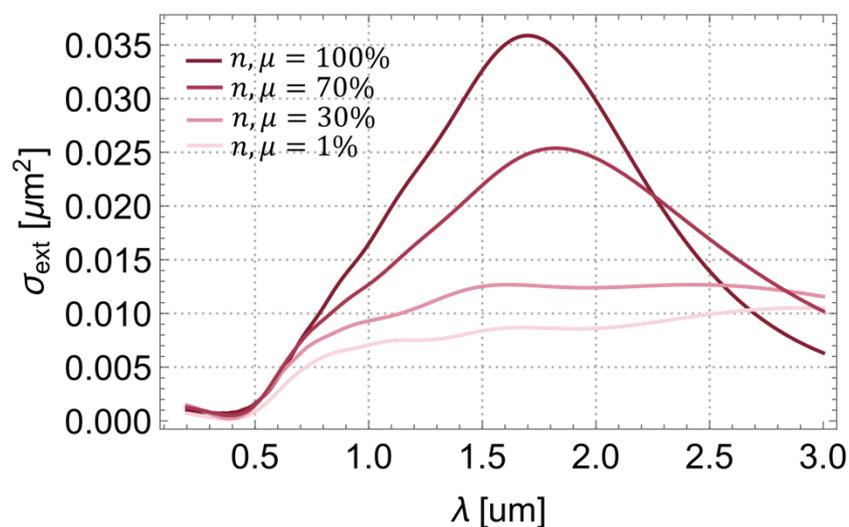

**Supplementary Fig. 16 | Simulations of extinction spectra with varying *n* and μ considering only the Drude-term in the Drude-Lorentz equation.** Extinction vs. wavelength for the same simulations as for the presented near-field profiles in Fig. 4e-h.

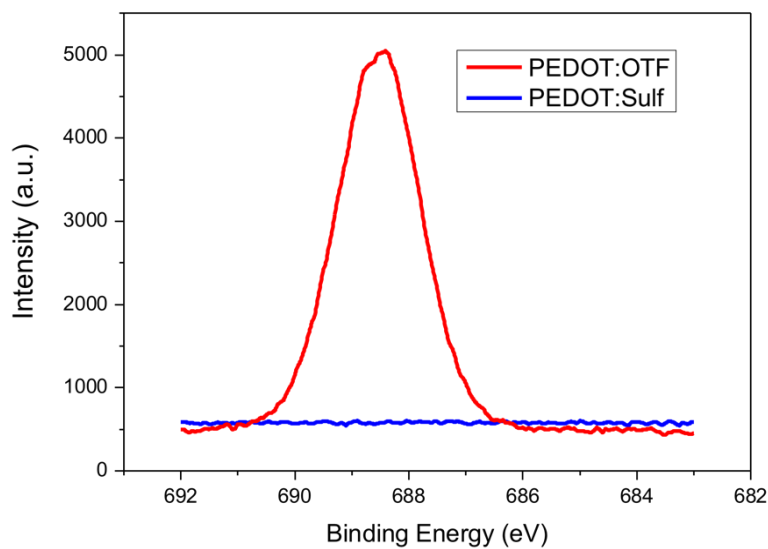

**Supplementary Fig. 17 | X-ray photoelectron spectroscopy for PEDOT thin films before and after acid treatment.** F1s core level region X-ray photoelectron spectroscopy spectra for pristine PEDOT:OTf thin film (red line) and PEDOT:Sulf thin film (blue line, PEDOT:Sulf).

**Supplementary Table 3 | Electrical conductivity of PEDOT:OTF and PEDOT:Sulf thin films measured using the 4 point probe method.** Electrical conductivity of PEDOT:OTF films and PEDOT:Sulf films on 50 nm sputtered ITO substrates measured via the 4-point

probe method. The 4-point probe measurements were obtained from an average of 10 spots in 2 different films for each measurement. The electrical conductivity of 50 nm sputtered ITO ranged from was: 843-902 S/cm.

| Sample | Number of times for vapor phase polymerization | Electrical conductivity, σ [S/cm] |
|---|---|---|
| ITO/PEDOT:OTF | 1 × | 2067 ± 133 |
| ITO/PEDOT:Sulf | 1 × | 2844 ± 111 |
| ITO/PEDOT:OTF | 2 × | 1666 ± 108 |
| ITO/PEDOT:Sulf | 2 × | 2774 ± 104 |
| ITO/PEDOT:OTF | 3 × | 1736 ± 177 |
| ITO/PEDOT:Sulf | 3 × | 2740 ± 293 |

**References:**

1. Chen, S. *et al.* On the anomalous optical conductivity dispersion of electrically conducting polymers: ultra-wide spectral range ellipsometry combined with a Drude–Lorentz model. *J. Mater. Chem. C* **7**, 4350–4362 (2019).
2. Kim, D., Jang, H., Lee, S., Kim, B. J. & Kim, F. S. Solid-State Organic Electrolyte-Gated Transistors Based on Doping-Controlled Polymer Composites with a Confined Two-Dimensional Channel in Dry Conditions. *ACS Appl. Mater. Interfaces* **13**, 1065–1075 (2021).
3. Ashcroft, N. W. & Mermin, N. D. *Solid state physics.* (Saunders college, 1976).
4. Chen, S. *et al.* Conductive polymer nanoantennas for dynamic organic plasmonics. *Nat. Nanotechnol.* **15**, 35–40 (2020).